\documentclass[entropy,review,submit,moreauthors,pdftex,10pt,a4paper]{mdpi} 
\firstpage{1} 
\articlenumber{x}
\doinum{10.3390/------}
\pubvolume{xx}
\pubyear{2017}
\copyrightyear{2017}
\externaleditor{Academic Editor: name}
\history{Received: date; Accepted: date; Published: date}
\preto{\abstractkeywords}{\nolinenumbers}

\pdfoutput=1

\usepackage{amsfonts}
\usepackage{amssymb}
\usepackage{physics}
\usepackage{amsbsy}
\usepackage{amsmath}
\newcommand{\beq}{\begin{equation}}
\newcommand{\eeq}{\end{equation}}
\newcommand{\bse}{\begin{subequations}}
\newcommand{\ese}{\end{subequations}}
\newcommand{\bea}{\begin{eqnarray}}
\newcommand{\eea}{\end{eqnarray}}
\newcommand{\fabri}[1]{{\color{blue}\small\textit{[[FT: #1]]}}}

\newcommand{\oper}[1]{\hat{#1}}
\newcommand{\rom}[1]{{\rm #1}}
\Title{UNCERTAINTY RELATIONS FOR COARSE--GRAINED MEASUREMENTS
: AN OVERVIEW}

\Author{Fabricio Toscano $^{1*}$\orcidA{}, Daniel S. Tasca $^{2}$, {\L}ukasz Rudnicki $^{3,4}$  and Stephen P. Walborn $^{1}$}
\address{%
$^{1}$ \quad Instituto de F{\'i}sica, Universidade Federal do Rio de Janeiro, Caixa Postal 68528, Rio de Janeiro, RJ 21941-972, Brazil\\
$^{2}$ \quad Instituto de F{\'i}sica, Universidade Federal Fluminense, Niteroi, RJ 24210-346, Brazil\\
$^{3}$ \quad Max-Planck-Institut f{\"u}r die Physik des Lichts, Staudtstra{\ss}e 2, 91058 Erlangen, Germany\\
$^{4}$ \quad Center for Theoretical Physics, Polish Academy of Sciences, Aleja Lotnik{\'o}w 32/46, 02-668 Warsaw, Poland}

\corres{Correspondence: toscano@if.ufrj.br; Tel.: +55-21-3938-7477}

\abstract{Uncertainty relations involving complementary observables are one of the cornerstones of quantum mechanics.  Aside from their fundamental significance, they play an important role in practical applications, such as detection of quantum correlations and security requirements in quantum cryptography. In continuous variable systems, the   spectra
of the relevant observables form a continuum and this necessitates the coarse graining of measurements.  However, these coarse-grained observables do not necessarily obey the same uncertainty relations as the original ones, a fact that can lead to false results when considering applications.  That  is, one cannot naively replace the original observables in the  
uncertainty relation for the coarse-grained observables and expect consistent results.  As such, a number of uncertainty relations that are specifically designed for coarse-grained observables have been developed.  In recognition of the 90$^{th}$ anniversary of the seminal Heisenberg uncertainty relation, celebrated last year, and all the subsequent work since then, here we give a review of the state of the art of coarse-grained uncertainty relations in continuous variable quantum systems, as well as their applications to fundamental quantum physics and quantum information tasks. Our review is meant to be balanced in its content, since both theoretical considerations and experimental perspectives are put on an equal footing.}


\begin{document}


\section{Introduction}

The physics of classical waves distinguishes itself from that of a classical point particle in a number of ways.  Waves are spread-out packets of energy moving through a medium, while a particle is localized and follows a well-defined trajectory. 
It was thus most surprising when it was discovered in the early 20th century that quantum objects, such as electrons and atoms, could exhibit behavior that at times was best described according to wave mechanics.  Moreover, it was shown that either wave or particle behavior could be observed depending almost entirely upon how an observer chooses to measure the system.  This complementarity of wave and particle behavior played a key role in the early debates concerning the validity of quantum theory \cite{wheeler83}, and has been linked to a number of interesting and fundamental phenomena of quantum physics \cite{scully91,kim00,bertet01,walborn02a}.  Though a number of complementarity relations have been cast in quantitative forms \cite{mandel91,englert96}, perhaps complementarity is most frequently observed in terms of quantum uncertainty relations.  In words, uncertainty relations establish the fact that the intrinsic uncertainties associated to measurement outcomes of two complementary observations of a quantum system can never both be arbitrarily small.    We note that this type of behavior appears in classical wave mechanics, for example in the form of time-bandwidth uncertainty relations, which are quite important in communications and signal processing \cite{ozaktas01}.  In contrast, there is no aspect of a classical physics that prohibits us from measuring all of the relevant properties of a classical point particle, at least in principle.
\par
 In addition to quantum fundamentals, quantum uncertainty relations play an important role in a number of interesting tasks associated to quantum information protocols, such as the detection of quantum correlations  and the security of quantum cryptography \cite{Coles2017}.  In this paper we focus on continuous variable (CV) quantum systems \cite{braunstein05,Adesso2014}. Though many interesting results have been found for discrete systems, they are outside the scope of this manuscript.  We refer the interested reader to Ref. \cite{Coles2017}, being a comprehensive unification and extension of two older reviews on entropic uncertainty relations, more focused on the physical \cite{bialynicki11} and information-theoretic \cite{Wehner2010} side respectively. However, since the coarse-grained scenario situates itself somehow in-between the discrete and continuous description, we make a short introduction to discrete entropic uncertainty relations before discussing their coarse-grained relatives. 
\par
 In CV systems, one encounters a fundamental problem when performing measurements.  That is, the eigenspectra of the corresponding observables are infinite dimensional, and can be continuous or discrete.  Since any measurement device registers measurement outcomes with a finite precision and within a finite range of values, the experimental assessment of CV observables can be quite different from theory.  Of course, one can consider a  truncation of the relevant Hilbert space \cite{sperling09a}, as well as some type of binning or coarse graining of the measurement outcomes.  This is similar to the idea of coarse graining that was discussed by Gibbs \cite{gibbs} and used by Paul and Tanya Ehrenfest \cite{ehrenfest1,ehrenfest2} in the early 20th century to account for imprecise knowledge of dynamical variables in statistical mechanics \cite{mackeybook}.  Coarse graining has also appeared in the quantum mechanical context as an attempt to describe the quantum--to--classical transition, where the idea is that measurement imprecision could be responsible for the disappearance of quantum properties \cite{kofler07,kofler08,raeisi11,wang13,jeong14}.  Though this is quite an intuitive notion, it was recently shown that one can always find an uncertainty relation that is satisfied non-trivially for any amount of coarse graining \cite{rudnicki12b}.  That is, quantum mechanical uncertainty is always present in this type of ``classical" limit.  This motivates the formulation of coarse-grained uncertainty relations.        
\par
In addition to the necessity of coarse graining, there could be practical advantages:  for tasks such as entanglement detection, it might be interesting to perform as few measurements as possible, advocating the use of coarse-grained measurements.  However, improper handling of coarse graining can result in false detections of entanglement \cite{Ray13b,Tasca13}, pseudo-violation of Bell's inequalities or the Tsirelson bound \cite{tasca09b,semenov11}, and sacrifice security in quantum key distribution \cite{Ray13a}, for example.   Thus, the proper formulation and application of uncertainty relations for coarse-grained observables is both interesting and necessary.            
\par
In the present contribution we review the current state of the art of uncertainty relations (URs) for coarse-grained observables in continuous-variable quantum systems.    In section \ref{sec:URs} we review the concept of uncertainty of continuous variable (CV) quantum systems in more depth and introduce several prominent URs.  In section \ref{sec:relevance} we discuss the utility of CV URs in quantum physics and quantum information, in particular for identifying non-classical states and quantum correlations.  Section \ref{Section:realistic-CG-models} presents the problem of coarse graining of CVs in detail, and two coarse-graining models are provided.  The current status of URs for these coarse-graining models is reviewed in section \ref{sec:CGUR}, where we present a series of coarse-grained URs previously reported in the literature \cite{bialynicki84,bialynicki06,bialynicki11,rudnicki12a,rudnicki12b}.  In addition, we extend the validity of some of these URs to general linear combinations of canonical observables. Section \ref{sec:appCGUR} is devoted to the experimental investigation and application of coarse-grained URs in quantum physics and quantum information.  Concluding remarks are provided in section \ref{sec:conc}.

\section{Uncertainty relations}
\label{sec:URs}
The history of uncertainty relations traces back to the early days of the formalization of quantum theory
and begins with the celebrated work by Heisenberg in 1927 \cite{Heisenberg1927} (see  \cite{wheeler83} for an English version). The work discussed what later became known as Heisenberg's uncertainty principle.  The first mathematical formulation for this principle, in \cite{Heisenberg1927}, essentially reads:
\beq
\label{Hrelation}
\Delta x\Delta p \gtrsim h
\eeq
where $\Delta x$ and $\Delta p$ are the uncertainties of the position and linear momentum of a particle, respectively, and $h$ is the Planck constant. Although the existence of such a principle is ultimately due to the non-commutativity of the position and momentum observables, it took almost 80 years for 
all the physical meanings, scope and validity of this principle to be elucidated \cite{Busch2007}.
Distinct physical meanings emerge from different definitions for ``uncertainty''
of position or momentum, and in each case a proper
multiplicative constant makes the lower bound sharp.
All of these inequalities are known by the generic name of {\it Uncertainty Relations}, from the beginning of this review referred to as URs. 
Even though the inception of the URs was made in the context of position and 
momentum of a particle, their existence can be extended to the ``uncertainties'' associated 
with any pair of non-commuting observables in discrete or continuous variable quantum systems.  Thus, generically we can define the URs as inequalities that stem from the fact that the measured quantities involved are associated to non-commuting observables.
\par 
Nowadays, we can say that it is clear that there are three conceptually distinct types of 
URs \cite{Busch2007}: {\it i)}  URs associated with the statistics of 
the measurement results of non-commuting observables after preparing the system
repeatedly  in the same quantum state, or {\it statistical} URs for short,  {\it ii)} the {\it error-disturbance} URs, also known as {\it noise-disturbance} URs, for the relation of the imprecision 
in the measurement of one observable and the corresponding disturbance in the other,  and, {\it iii)} the {\it joint measurement} URs associated with the precision of the joint measurements of non-commuting observables.
The {\it error-disturbance} URs has two main contributions: one in 
Refs. \cite{Ozawa2003,Ozawa2004,Ozawa2005} and the other in Refs. \cite{Werner2004,Busch2004,Busch2013}. There was a certain controversy involving these two contributions, drawn by their individual claims to follow the original truth of Heisenberg's ideas. The conclusion 
of this controversy is that if you define measures of error and disturbance for an individual state, then the UR for these measures is not given by  
Eq.\eqref{Hrelation} \cite{Ozawa2004}. However, if one gives a state-independent characterisation of the overall performance of measuring devices as a measure of uncertainty, then an UR of the form given in Eq.\eqref{Hrelation} applies \cite{Busch2013}.
The development of {\it joint measurement} URs has an early contribution in Ref.
\cite{Arthurs1965} and further developments were given in Refs. \cite{Arthurs1988,Ishikawa1991,Raymer1994,Ozawa2004a}.
\par
The {\it statistical} URs are also referred to in the literature as 
{\it preparation} URs. This is because it is impossible to prepare a quantum system in a state for which two non-commuting observables have sharply defined values.
However, here we prefer to call them {\it statistical} URs, 
as they express the limits to the amount of information that can be obtained about complementary properties of a quantum system when it is repeatedly measured after 
being prepared in the same initial state in each round of the measurement 
process. We emphasize that there is not any attempt to measure the 
two commuting observables simultaneously.  In each round of the measurement process
only one observable is measured, the choice of which could be made randomly.
In this sense the "uncertainties" contained in the {\it statistical} URs are of the 
statistical type: the more certain the sequence of outcomes of one observable is in a given state, then the more uncertain is the sequence of outcomes of the other non-commuting observable(s) considered. 
\par
This review focuses on  {\it statistical} URs that are valid for coarse-grained 
measurements in continuous variable quantum systems, although a similar approach can be made for the other two types of URs mentioned above.
Here, we follow modern Quantum Information Theory (QIT) that classifies 
physical systems according to the type of quantum states accessible to them. 
According to this catalogue there are two main types of quantum systems: those where 
the Hilbert space of quantum states has finite dimension and those where it has infinite dimension.  In particular, we are interested only in continuous variable (CV) systems where the Hilbert 
space, ${\cal H}$, of pure states, $\ket{\psi}$, has an infinite dimension. 
The CV systems that we consider consist of a finite set of $n$ bosonic modes, sometimes called ''qumodes'' \cite{braunstein05}, so that ${\cal H}:={\cal H}_1\otimes\ldots\otimes {\cal H}_n$. Each mode is described 
by a pair of canonically conjugate operators, $\hat x_j$ and $\hat p_j$, such that 
\beq 
\label{comutatorxp}
[\hat x_j,\hat p_k]=i\hbar \hat 1\delta_{jk}  .
\eeq 
Alternatively, each mode can be described by a pair of ladder operators,
$\hat a_j:=(1/\sqrt{2\hbar})(\hat x_j+i\hat p_j)$ and 
$\hat a_j^\dagger:=(1/\sqrt{2\hbar})(\hat x_j-i\hat p_j)$, with $[\hat a_j,\hat a_k^\dagger ]=\hat 1\delta_{jk}$. Therefore, the separable Hilbert space of each mode, ${\cal H}_j$, has a enumerable basis $\{\ket{n_j}\}_{n_j=1,\ldots,\infty}$ consisting of eigenstates of the number operator, {\it viz.} $\hat n_j\ket{n_j}=n_j\ket{n_j}$, evidencing the 
infinite dimensionality of the Hilbert space of the quantum states.
In the case of mixed states we use density operators represented by greek letters with a hat, {\it i.e}  $\hat \rho$, $\hat \sigma$ etc. 

Important examples of CV systems are the motional degrees of freedom of atoms, ions and molecules, where $\hat x_j$ and $\hat p_j$ are the components of the position and linear momentum of the particles \footnote{In this case 
 $\hbar$ in Eq.(\ref{comutatorxp}) is the usual reduced Planck constant, {\it i.e.} $\hbar=h/2\pi$.}; the quadrature modes of the quantized electromagnetic field where 
$\hat x_j$ and $\hat p_j$  are canonically conjugate quadratures \footnote{In this case $\hbar$ in  Eq.(\ref{comutatorxp}) is just $\hbar=1$. } \cite{braunstein05}; and the transverse spatial degrees of freedom of single photons propagating in the paraxial approximation 
\footnote{In this case $\hbar$ in Eq.(\ref{comutatorxp}) is $\hbar=\lambda/2\pi$ where $\lambda$ is the photon's wave length.} \cite{tasca11}.
\par
In what follows we summarize the principal {\it statistical} URs in CV systems that have been generalised to coarse-grained measurements. The corresponding coarse-grained URs will be presented in Section \ref{sec:CGUR}.

\subsection{Heisenberg (or variance) Uncertainty Relation}
\label{sec:varianceURs}
Let us consider two operators:
\beq
\label{defuv}
\hat u:={\bf d}^T{\bf \hat  x}={\bf a}^T{\bf\hat q}+{\bf a}^{\prime T}{\bf \hat p}\;\;\mbox{and}\;\;
\hat v:={\bf d}^{\prime T}{\bf \hat  x}={\bf b}^T {\bf\hat q}+{\bf b}^{\prime T}{\bf \hat p},
\eeq
where $T$ means transposition and we define the $2n$-dimensional vector of operators,
\beq
{\bf \hat x}:=({\bf\hat q},{\bf\hat p})^T=(\hat x_1,\ldots,\hat x_n,\hat p_1,\ldots,\hat p_n)^T,
\eeq
as well as the arbitrary real vectors,
\beq
\label{defofd}
{\bf d}=({\bf a},{\bf a}^\prime)^T=(a_1,\ldots,a_n,a^\prime_1,\ldots,a_n^\prime)^T\;\;\;\mbox{and}\;\;\;
{\bf d}^\prime=({\bf b},{\bf b}^\prime)^T=(b_1,\ldots,b_n,b^\prime_1,\ldots,b_n^\prime)^T.
\eeq
The commutation relation of $\hat u$ and $\hat v$ is 
\beq
\label{CCR}
[\hat u,\hat v]=i\hbar {\bf d}^T\mathbf{J}{\bf d}^\prime\hat{1}=:i\hbar \gamma\hat{1},
\eeq
where $\mathbf{J}$ is the $2n \times 2n$-dimensional matrix of the symplectic norm
\cite{Dutta1995}:
\beq
\mathbf{J}=   \begin{pmatrix} 
      \mathbf {O} &\mathbf{I} \\
    -  \mathbf {I} & \mathbf{O} \\
   \end{pmatrix},
\eeq
and the $n\times n$ matrices in the blocks are the identity matrix $\mathbf I$
and the null matrix $\mathbf O$. In this review, matrices of an arbitrary shape not treated as quantum-mechanical operators are denoted in bold and without a hat. 
\par
The parameter $\gamma$ in definition \eqref{CCR} is a scalar that in some sense quantifies the non-commutativity of $\hat u$ and $\hat v$.
Commutation relations such as Eq.(\ref{CCR}) are called Canonical Commutation Relations (CCR) \footnote{Sometimes the name CCR is used in the 
case when $\gamma=1$, however, as $\hbar \gamma $ can be interpreted as an effective Planck constant, so the name CCR here is well justified.}. However, a CCR between two operators $\hat u$ and $\hat v$ 
does not guarantee that they are necessarily Canonically Conjugate Operators (CCOs). 
For this to be true we additionally need that the eigenvectors of $\hat u$ and $\hat v$
must be connected by a Fourier Transform. In such a case we call $\hat u$ and $\hat v$
CCOs \footnote{Also note that when two operators like the ones defined in Eq.(\ref{defuv}) have their eigenstates connected by a Fourier Transform, they necessary satisfy a commutation relation like in Eq.(\ref{CCR}), as can be easily shown. However the converse is not true. Take for example the single mode operators $\hat u=\hat x$ and $\hat v=\hat x+\hat p$, which satisfy $[\hat u,\hat v]=[\hat x,\hat p]=i\hbar$ but are not a Fourier pair.}.
\par
Every pair of operators, $\hat u$ and $\hat v$, that obey a CCR also satisfies the
 {\it statistical} UR:
\beq
\label{VarianceUR}
\sigma_{P_u}^2\sigma_{P_v}^2
\ge \frac{\hbar^2}{4}\gamma^2,
\eeq
where 
\beq
\label{defvariance}
\sigma_{P_u}^2:=\langle\hat u^2\rangle-\expval{\hat u}^2,\;\;\;\mbox{and}\;\;\;
\sigma_{P_v}^2:=\langle\hat v^2\rangle-\expval{\hat v}^2,
\eeq
are the variances of the marginal probability distribution functions (pdf):
\beq
\label{marginalPDuv}
P_{u}(u)=\expval{\dyad{u}},\;\;\;\mbox{and}\;\;\;
P_{v}(v)=\expval{\dyad{v}},
\eeq
where we have defined
\beq
\expval{\ldots}:=\Tr(\ldots \hat \rho),
\eeq
with $\hat \rho$ being an arbitrary $n-$mode quantum state.
We call the UR in Eq.(\ref{VarianceUR}) the {\it Heisenberg} UR, or variance-product UR. 
For one mode CCOs, such as $\hat u=\hat x$ and $\hat v=\hat p$ (therefore $\gamma=1$),
the {\it Heisenberg} UR in Eq.(\ref{VarianceUR}) was first proved by Kennard in 1927 \cite{Kennard1927},
inspired by the inequality in Eq.(\ref{Hrelation}) of Heisenberg's seminal paper of the same year \cite{Heisenberg1927}. Later, it was also proved by Weyl in 1928 \cite{Weyl1928}. 
In 1929 Robertson  \cite{Robertson1929}  extended the {\it Heisenberg} UR for any pair 
of Hermitian operators $\hat A$ and $\hat B$:
\beq
\label{RobertsonUR}
\sigma_{P_A}^2\sigma_{P_B}^2\geq \frac{1}{4}\left|\expval{[\hat A,\hat B]}\right|^2.
\eeq
This result extends the {\it Heisenberg } UR in Eq.(\ref{VarianceUR}) to
$\hat u$ and $\hat v$ that are not CCOs. 

For every
variance-product UR in Eq.(\ref{RobertsonUR}) there is an associated linear UR:
\beq
\label{linearUR}
\sigma_{P_A}^2+\sigma_{P_B}^2\geq \left|\expval{[\hat A,\hat B]}\right|.
\eeq
In fact, this UR is a consequence of Eq.(\ref{RobertsonUR}) and the trivial inequality $(\sigma_{P_A}-\sigma_{P_B})^2\ge 0$, so that
\beq
\label{chainHURlinear}
\sigma_{P_A}^2+\sigma_{P_B}^2\geq 2\sigma_{P_A}\sigma_{P_B}\geq \left|\expval{[\hat A,\hat B]}\right|, 
\eeq
where it also follows that the linear UR is weaker than the variance product UR.
In 1930 Schr\"odinger \cite{Schrodinger1930} improved the lower bound in Eq.(\ref{RobertsonUR}),  so the new stronger UR reads:
\beq
\label{SchrodingerUR}
\sigma_{P_A}^2\sigma_{P_B}^2\geq \frac{1}{4}\left|\expval{[\hat A,\hat B]}\right|^2
+\frac{1}{4}\left|\expval{\{\hat A-\expval{\hat A},\hat B-\expval{\hat B}\}}\right|^2,
\eeq
where $\{ , \}$ is the anti-commutator.
\par 
One  interesting property of the {\it Heisenberg} UR in Eq.(\ref{VarianceUR}) is that
the lower bound is independent of the quantum state $\hat \rho$ under consideration.  
Another property
is that it can be seen as a {\it bona fide} condition on the covariance matrix
of an $n-$mode quantum state $\hat \rho$, {\it viz} the matrix of second moments 
of the CCOs, contained in the vector ${\bf \hat x}$, of the state $\hat \rho$:
\beq
\mathbf{V}:=\frac{\expval{{\bf \hat x}{\bf \hat x}^T}+\expval{{\bf \hat x}{\bf \hat x}^T}^T}{2}-\expval{\bf \hat x}\langle{\bf \hat x}^T\rangle.
\eeq
Indeed, in \cite{Simon1994,Solomon2012} it was shown 
that  the {\it bona fide} condition on the covariance matrix $\mathbf{V}$ of a quantum state $\hat \rho$ is,
\beq
\label{bonafideC}
\mathbf{V}+\frac{i\hbar}{2}\mathbf{J} \ge 0,
\eeq
where the inequality means that the matrix on the left hand side is positive semi-definite, {\it viz.} all of
its eigenvalues are greater or equal to zero. 
Applying the inequality in Eq.(\ref{SchrodingerUR}) to the canonical conjugate 
operators $\hat x$ and $\hat p$, we have,
\beq
\label{detVUR}
\sqrt{\det(\mathbf{V})}=\sqrt{\sigma_{P_x}^2\sigma_{P_p}^2-\frac{1}{4}
\left|\expval{\{\hat x-\expval{\hat x},\hat p-\expval{\hat p}\}}\right|}\ge \frac{\hbar}{2}.
\eeq
For one mode systems this inequality is equivalent to the {\it bona fide} condition
in Eq.(\ref{bonafideC}). However, for multimode systems it is not enough.
For multimode systems, a way to verify the {\it bona fide} of the covariance matrix was 
given in
\cite{simon00,huang11}, where it was shown 
 that to verify the condition in Eq.(\ref{bonafideC}) is equivalent to verify the linear UR in Eq.(\ref{linearUR}) for all the operators, $\hat u$ and $\hat v$, defined in Eq.(\ref{defuv}).
Therefore, using Eq.(\ref{chainHURlinear}) we can write the series of implications:
\beq
\sigma_{P_u}^2\sigma_{P_v}^2
\ge \frac{\hbar^2}{4}\gamma^2\;\;\Rightarrow
\sigma_{P_u}^2+\sigma_{P_v}^2
\ge \hbar|\gamma|\Leftrightarrow
\mathbf{V}+\frac{i\hbar}{2}\mathbf{J} \ge 0.
\eeq
Thus, it is enough to verify the violation of the {\it Heisenberg} UR for some pair of operators
$\hat u$ and $\hat v$ to confirm that the {\it bona fide} condition on the covariance matrix of some  
$n-$mode operator $\hat \rho$ is not satisfied.  

\subsection{Entropic URs}
\label{EntropicUR}
The use of entropy functions to quantify uncertainty of a probabilistic variable dates back to the early work of Shannon \cite{shannon48}.  Since then, a number of different entropy functions have been defined, with distinct relations to meaningful characteristics of the probability distributions considered.  A number of these entropy functions have found use in quantum mechanics and, in particular, in QIT \cite{Coles2017}.  Here we outline the application of these functions to uncertainty relations between non-commuting observables.

\subsubsection{\it Shannon-entropy UR}

The UR based on the differential Shannon entropy for operators defined in Eq.(\ref{defuv})
is:
\beq
\label{ShannonUR}
h[P_u]+h[P_v]\geq \ln(\pi e \hbar |\gamma| ),
\eeq
where $P_u$ and $P_v$ are the marginal pdf
defined in Eq.(\ref{marginalPDuv}) and the differential Shannon entropy of a pdf, $P$, is defined as \cite{cover}:
\beq
h[P]:=-\int_{-\infty}^\infty dy\, P(y) \, \ln P(y).
\label{defShannonEntropy}
\eeq
For $\hat u$ and $\hat v$ as CCOs, this uncertainty relation was first proved in 1975 by Bialynicki-Birula \& Mycielski \cite{bialynicki75}. In their derivation the authors used the $L_p$-$L_q$ norm inequality for the Fourier transform operator obtained by Beckner \cite{beckner1975}. Note that in the literature this inequality is sometimes referred to as the Babenko-Beckner inequality\footnote{Eq. 1.104 from \cite{bialynicki11} provides an extension of this inequality to the case of arbitrary mixed states, using two variants of the Minkowski inequality.}, because Babenko \cite{babenko61} had proved it before Beckner, but only for certain combinations of $(p,q)$ parameters. For the sake of completeness, we should also mention that Hirschman \cite{hirschman1957} had derived a weaker version of \eqref{ShannonUR} with the constant $e \pi$ inside the logarithm replaced by $2\pi$.
The extension of the validity for operators  
$\hat u$ and $\hat v$ that are not CCOs was provided very recently in Refs.  
\cite{guanlei09,huang11}. 
\par
The {\it Shannon-entropy} UR is in general stronger than the {\it Heisenberg} UR
as the former implies the latter. This can be seen by using the 
 inequality for a pdf $P$ \cite{cover}: 
 \beq
 \label{prodvarlessentro}
 \ln(2\pi e \sigma_P^2)\ge 2h[P],
 \eeq
 where $\sigma^2$ is the variance of $P$. Therefore, we can write the chain of inequalities: 
 \beq
 \label{chainURs}
 \ln(2\pi e \sigma_{P_u}\sigma_{P_v})\ge h[P_u]+h[P_v]\ge \ln(\pi e \hbar |\gamma|),
 \eeq
 that compress the URs in Eqs.(\ref{VarianceUR}) and (\ref{ShannonUR}).
 It is clear from Eq.(\ref{chainURs}) that the verification of the {\it Shannon-entropy}
 UR for any pair of the operators in Eq.(\ref{defuv}) is enough to guarantee the {\it bona fide} condition in Eq.(\ref{bonafideC}) \cite{huang11}.
\par
When the quantum state $\hat \rho$ is Gaussian, {\it viz} when the Wigner function of 
 $\hat \rho$ is a multivariate Gaussian probability distribution \cite{Adesso2014}, 
 the marginal pdfs, $P_u$ and $P_v$, are also Gaussians.
 Remembering that the differential Shannon entropy of a Gaussian pdf
 $P$, with variance $\sigma_P^2$, is $h[P]=(1/2)\ln(2\pi e \sigma_P^2)$
 \cite{cover}, we can see that Gaussian states saturate the first inequality in Eq.(\ref{chainURs}). Therefore, for Gaussian states the {\it Heisenberg} UR and the 
 {\it Shannon-entropy} UR are completely equivalent. 
 As we will see in Section \ref{sec:CGUR} this is not the case for the coarse-grained versions of these URs.
\par  
\subsubsection{\it R\'enyi-entropy URs}
\par
The UR based on the differential R\'enyi entropy for the operators defined in Eq.(\ref{defuv}) that are CCOs is given by the inequality:
\beq
\label{RenyiUR}
h_\alpha[P_u]+h_\beta[P_v]\geq 
\ln\left(\frac{\pi \hbar |\gamma|}{\alpha^{\frac{1}{(2-2\alpha)}}\beta^{\frac{1}{(2-2\beta)}}}\right),
\eeq
where $1/\alpha+1/\beta=2$ with $1/2\le\alpha\le 1$ and $\gamma=1$ since we deal with CCO operators. As before, $P_u$ and $P_v$ are the marginal pdfs
defined in Eq.(\ref{marginalPDuv}) and the differential R\'enyi entropy of order $\alpha$ relevant for an arbitrary pdf, $P$, is defined as \cite{cover}:
\beq
\label{RenyiEntropy}
h_\alpha[P]:=\frac{1}{1-\alpha}\ln\left(\int_{-\infty}^\infty dy\; \left[P(y)\right]^\alpha\right).
\eeq
The  {\it R\'enyi-entropy} UR was proved recently (in 2006) by Bialynicki-Birula \cite{bialynicki06} (see also \cite{bialynicki11}) again with the help of the powerful mathematical tools developed in \cite{beckner1975}.
Note that in the limit $\alpha\rightarrow 1$ we also  have $\beta \rightarrow 1$, and consequently $\alpha^{\frac{1}{(2-2\alpha)}}\beta^{\frac{1}{(2-2\beta)}}\rightarrow 1/e$.
Therefore, in the limit $\alpha\rightarrow 1$ we have $h_\alpha[P_u]\rightarrow h[P_u]$ and $h_\beta[P_v]\rightarrow h[P_v]$, 
so the expression in Eq.(\ref{RenyiUR}) reduces to the {\it Shannon-entropy} UR  
in Eq.(\ref{ShannonUR}) for $\gamma=1$.
As far as we know, in contrast to the {\it Shannon-entropy} UR, the extension of the 
{\it R\'enyi-entropy} UR to the general case of operators that are not necessarily CCOs is still a challenge for the future.   A first attempt in this direction
was provided in Ref. \cite{guanlei09},
where the authors show  that the {\it R\'enyi } UR in Eq.\eqref{RenyiUR} is still valid 
when the eigenvectors of $\hat u$ and $\hat v$
are connected by a Fractional Fourier Transform \cite{ozaktas01},  which corresponds to rotation in phase space.
\par
All of the URs mentioned in this section (this is a general pattern though) can be cast in a general form
\beq
F(\hat \rho;\hat u, \hat v; P_u, P_v) \geq f(\hbar |\gamma|), 
\eeq
where $F$ is an uncertainty functional [left hand side of inequalities \eqref{VarianceUR}, \eqref{ShannonUR}, \eqref{RenyiUR} for example] and $f$ represents its respective lower bounds. In particular, we do not pay much attention here to the Tsallis entropy and URs associated with it. Again such URs can be cast in the general form stated above and their derivation is usually very similar in spirit to the case of the R{\'e}nyi entropy.
\par
In Section \ref{sec:relevance} we will summarise the relevance of the 
{\it statistical} UR in general and in particular 
the URs in Eqs.(\ref{VarianceUR}), (\ref{ShannonUR}) and (\ref{RenyiUR}).
In Section \ref{sec:CGUR} we will present versions of the {\it Heisenberg}, {\it Shannon-entropy} and {\it R\'enyi-entropy} URs for coarse-grained measurements.

\section{Utility of Uncertainty Relations in Quantum Physics}
\label{sec:relevance}
\par
Uncertainty relations can be applied in a number of useful and interesting ways.  First, they provide a way to test if experimental data 
is compatible with quantum mechanics.  This is particularly helpful in testing the experimental reconstruction of density matrices or phase-space distributions (quantum state tomography) or for example the covariance matrix \cite{narcowich90}, or any other set of moments of the CCOs of the modes.   
\par
URs can also be used to characterize non-classical states of light, such as squeezed states \cite{slusher85}.  In this case observation of the variance $\sigma_{P_u}^2  \leq \hbar/4$ where $\hat u$ is a phase-space quadrature
in Eq.(\ref{defuv}), indicates noise fluctuations in this quadrature that are smaller than the vacuum state.  As a consequence of the Heisenberg UR, the noise fluctuations in the conjugate quadrature must be larger or equal to $\hbar/4 \sigma_{P_u}^2$.  In a similar fashion, in Ref. \cite{shchukin05c} it was shown that violation of one out of an infinite hierarchy of inequalities involving normally ordered quadrature moments is sufficient to demonstrate non-classicality.  We note that $\sigma_{P_u}^2  \leq \hbar/4$ corresponds to the lowest-order inequality of this set.  Related techniques have been developed based on the quantum version of Bochner's theorem for the existence of a positive semi-definite characteristic function \cite{vogel00,richter02}.  Both of these methods have been used experimentally in Ref. \cite{kiesel09}.  More recently, these two techniques were unified into a single criteria involving derivatives of the characteristic function \cite{ryl15}, and put to test on a squeezed vacuum state.           
\par
To our knowledge, the first application of URs to identify quantum correlations was described in Ref. \cite{reid88}, in which the authors proposed a  Heisenberg-like UR, similar to that in Eq.(\ref{VarianceUR}), to identify non-classical correlations between both the phases and intensities of the fields produced by a non-degenerate parametric oscillator.  It was shown by M. Reid \cite{reid89} that these measurements provide a method to demonstrate correlations for which the seminal Einstein-Podolsky-Rosen (EPR) argument \cite{epr35} is valid.  An experiment using this UR-based method to demonstrate EPR-correlations between light fields was realized shortly therafter \cite{ou92}. It was later shown by Wiseman et al. \cite{wiseman07,jones07} that the Reid EPR-criterion was indeed a method to identify quantum states  that violate a ``local hidden state" model of correlations.  This type of  correlation has been called ``EPR-steering", or just ``steering" \cite{cavalcanti17}, as this was the terminology used by Schr\"odinger when he discussed EPR correlations in 1935 \cite{schrodinger35}.  Since 2007, EPR-steering has been understood to make up part of a hierarchy of quantum correlations, situated between entanglement \cite{horodecki09,guhne09} and Bell non-locality \cite{brunner14}.   In addition to methods utilizing variance-based URs \cite{ji15}, entropic URs, such as those in Section \ref{EntropicUR}, can be used to identify EPR-steering \cite{walborn11a,schneeloch13}.  Some of these URs can be used to test security in continuous variable quantum cryptography \cite{reid00,grosshans04}, and it has been shown that violation of entropic EPR-steering criteria are directly related to the secret key rate in one-sided device independent cryptography \cite{branciard12}.  We also highlight techniques based on a matrix--of--moments approach \cite{kogias15}. Continuous-variable EPR-steering has been observed in intense fields \cite{ou92,silberhorn01,bowen03} as well as photon pairs \cite{dangelo04,howell04,tasca09a,walborn11a}.    
\par
Perhaps one of the most important tasks in quantum information is identifying quantum entanglement. In this respect, URs have also found widespread use in simple and experimentally friendly entanglement detection methods, as we will now describe.  Several early entanglement criteria for bipartite CV systems were developed using URs  \cite{duan00,mancini02,giovannetti03,zhang10}.  A particularly convenient method to construct entanglement criteria is to use the Peres-Horedecki positive partial transposition argument \cite{peres96,horodecki96} (PPT), and apply it to uncertainty relations \cite{nha08,walborn09,guhne09,saboia11,toscano15}.  The PPT argument is as follows.  A bipartite separable state $\hat  \sigma_{12}$ can be written as \cite{werner89}
\begin{equation}
\hat \sigma_{12} = \sum_i \lambda_i \hat \rho_{1i} \otimes \hat \rho_{2i}, 
\label{eq:sep}
\end{equation} 
where $\hat \rho_{1i}$ and $\hat \rho_{2i}$ are {\it bona fide} density operators of subsystems 1 and 2, respectively. 
The transpose of the state $\hat \rho_{2i}$, here denoted $\hat \rho_{2i}^{T}$, is still a positive operator, since full transposition preserves the eigenspectrum.  Thus, partial transposition (with respect to second subsystem) of $\hat \sigma_{12}$ gives the valid quantum state:
\begin{equation}
\hat \sigma_{12}^{T_2} = \sum_i \lambda_i \hat \rho_{1i} \otimes \hat \rho_{2i}^{T}. 
\end{equation} 
On the other hand, partial transposition of an entangled state $\hat \varrho_{12}$, which cannot be written in the form \eqref{eq:sep}, can lead to a non-physical density matrix since partial transposition may not preserve the positivity of the eigenspectrum.   Thus, one can identify entanglement in a bipartite density operator by calculating the partial transposition and searching for negative eigenvalues, and even quantify the amount of entanglement via the negativity \cite{vidal02}. However, applications of this method in experiments requires quantum state tomography and reconstruction of  the density operator, which involves a large number of measurements. A more experimentally friendly method to identify entanglement is to evaluate an UR applied to the partial transposition of $\hat \varrho_{12}$, which we describe in the next paragraph.  The PPT-argument is only a sufficient entanglement criteria in a general bipartition of $m\times(n-m)$ modes, but is necessary and sufficient in the particular case of bipartitions of $1 \times (n-1)$ modes in CV Gaussian states \cite{werner01,braunstein05}. Thus, there are no Gaussian states which are PPT entangled states in bipartitions 
of the form $1 \times (n-1)$. However, there do exist entangled CV Gaussian states that are PPT in general bipartitions of the type  $m\times(n-m)$. These are called  bound entangled states \cite{horodecki98}. In Gaussian states, this set of bound entangled states coincides with the set of all states whose entanglement
in a bipartition $m\times(n-m)$ cannot be distilled using local operations and classical communication \cite{bennett96,giedke01a,giedke01b}. However, to our knowledge, for non-Gaussian states it is conjectured that 
the set of bound entangled states in a given bipartition  is only a sub-set of the set 
of undistillable states in that bipartition.  
\par
For continuous variables, R. Simon showed that transposition is equivalent to a momentum reflection, taking the single mode Wigner phase-space distribution $\mathcal{W}({\bf x},{\bf p}) \longrightarrow \mathcal{W}^{T_2}({\bf x},{ \bf p})=\mathcal{W}({\bf x}, \mathbf{T} {\bf p})$ \cite{simon00}, where $\mathbf{T}$ is a diagonal matrix whose elements are $+1$ for non-transposed modes, and $-1$ for the transposed ones.  Thus, evaluating the "transposed" Wigner function is the same as evaluating the original Wigner function with a sign change in the reflected $p$ variables.
\par
For simplicity, we consider now the particular example of global operators of a bipartite state:
\begin{equation}
\oper{u}_{\pm} = \oper{u}_1 \pm \oper{u}_2,
\label{eq:upm}
\end{equation}
and    
\begin{equation}
\oper{v}_{\pm} = \oper{v}_1 \pm \oper{v}_2.
\label{eq:vpm}
\end{equation} 
We note that operators with the same sign satisfy the commutation relations $[\oper{u}_{\pm},\oper{v}_{\pm}]=2 i \hbar \gamma$ , so that these non-commuting operators after being an input to the uncertainty functionals fulfill the UR of the aforementioned form [note the factor of $2$ in the argument of $f(\cdot)$]
\begin{equation}
F(\hat \varrho_{12};\hat u_\pm, \hat v_\pm; P_{u_{\pm}},P_{v_{\pm}})  \geq f(2\hbar |\gamma|).
\label{eq:FUR}
\end{equation}  
 Using the transformation of the Wigner function under partial transposition described above, one can evaluate the uncertainty functional of the partially transposed state $\hat \varrho_{12}^{T_2}$ via measurements on the actual state $\varrho_{12}$ using the relation
  \begin{equation}
F(\varrho_{12}^{T_2}; \hat u_\pm, \hat v_\pm; P_{{u}_{\pm}},P_{v_{\pm}}) = F(\varrho_{12};\hat u_\pm, \hat v_\mp; P_{{u}_{\pm}},P_{v_{\mp}}), 
\end{equation}  
which can be lower than $ f(2\hbar |\gamma|)$ since the operators with different signs do commute. 
This possibility, when experimentally confirmed, indicates that $\varrho_{12}^{T_2}$ is not a {\it bona fide} density operator, and thus the bipartite quantum state $\varrho_{12}$ is  entangled.

Building on this general reasoning (PPT argument applied to an UR) several entanglement criteria have been developed. A comprehensive list of the criteria contains those based on the variances \cite{hyllus06,nha07} and higher-order moments \cite{agarwal05,hillery06a}, Shannon entropy \cite{walborn09}, R\'enyi entropy \cite{saboia11}, characteristic function \cite{paul18b} as well as the triple product variance relation \cite{paul16}.  Particularly powerful is the formalism developed by Shchukin and Vogel, which provides an infinite set of inequalities involving moments of the bipartite state \cite{shchukin05}, such that violation of a single inequality indicates entanglement.  We note that some of these criteria can be applicable to any non-commuting global operators.  Uncertainty-based approaches (using the PPT method directly or not) have been developed for multipartite systems \cite{vanloock03,sun09}, and a general framework to construct entanglement criteria for multipartite systems based on the ''PPT+UR'' interrelation was presented in Ref. \cite{toscano15}. The Shchukin-Vogel hierarchy of moment inequalities has also been applied to the multipartite case \cite{shchukin06}.  
  \par
The PPT+UR approach has been used to identify continous variable entanglement experimentally in a number of systems, including entangled fields from parametric oscillators and amplifiers \cite{bowen03,villar05,coelho09} as well as spatially entangled photon pairs produced from parametric down conversion \cite{howell04,tasca08,paul16}, and time/frequency entangled photon pairs \cite{shalm12,maclean18}.  A higher-order inequality in the Shchukin-Vogel criteria \cite{shchukin05} has been used to observe genuine non-Gaussian entanglement \cite{gomes09b}.

\section{Realistic coarse-grained measurements of continuous distributions}
\label{Section:realistic-CG-models}

Coarse graining of observables with continuous spectra is a consequence of any realistic measurement process. In the laboratory, an experimentalist is given the task of designing projective measurements in order to recover information about probability densities of a continuous variable quantum system. Naturally, only partial information about the underlying continuous structure of the infinite-dimensional physical system is retrieved in a laboratory experiment. Whichever measurement design is chosen, the experimentalist is faced with two main difficulties, namely the finite {\it detector range} and finite {\it measurement resolution}, related to the size of the total region of possible detection events and the precision in which events are registered, respectively. The detector range problem \cite{Ray13a,Ray13b} results from the finite amount of resource available to the experimentalist. For instance, consider a position discriminator based on a multi-element detector array. The array has a spatial reach (in a single spatial dimension) that increases linearly with the number of detectors. In a similar fashion, the sampling time of a single element detector used in raster scanning mode increases linearly with the chosen detection range. Continuous variables such as the position are also inevitably affected by the inherent finite resolution of the measurement apparatus \cite{rudnicki12a}, such as the size of each individual detector in the array, or the pixel size of a camera.  Altogether, the finite detector range and measurement resolution restrict the capability to probe the detection position, limiting the experimentalist to a {\it coarse-grained} sample of the underlying CV degree of freedom.

The constraints imposed by the finite spatial reach and resolution of the measurement apparatus are then important features that must be considered in the experiment design. Ideally, the experimentalist would chose measurement settings producing the finest coarse-grained sample possible. As a trade-off, the increased resolution entails the sampling of a greater number of pixels (if the range of detection is preserved), increasing the amount of resources used in data acquisition and analysis. The compromise between the used resource and chosen resolution depends on the specific design and measurement technique. A single raster scanning detector is inherently inefficient and leads to acquisition times that grow with the number of scanned outcomes. On the other hand, the acquisition time is dramatically reduced by the use of multi-element detector arrays \cite{edgar12,aspden13,moreau14,tentrup17}. Other techniques such as position-to-time multiplexing \cite{warburton11,leach12b} allow the sampling of multiple position outcomes with single element detectors, but at the expense of an increased dead-time between consecutive detections.  We have exemplified the finite detector range and finite measurement resolution problems in terms of a detector that registers the position of a particle.  However, similar considerations are valid for any detection system that registers a digitalized value of a continuous physical parameter.
\par
Under constraints of resource utilisation --such as the number of detectors and/or sampling time-- the experimentalist needs to set the number of possible detection outcomes for their coarse-grained measurements. Therefore, a natural question that arises regards the coarse-graining design allowing the extraction of the desired information. Naively, one might think that usual quantum mechanical features learnt from physics textbooks would be directly observable from the coarse-grained distributions obtained in the laboratory. The most prominent counter-example is the experimental observation of the {\it Heisenberg} UR in Eq.(\ref{VarianceUR}).  As shown in Ref. \cite{rudnicki12a}, coarse-grained distributions of conjugate continuous variables do not necessarily satisfy the well known UR valid for continuous distributions. In order to accurately inspect the uncertainty product of the measured distributions in accordance with the {\it Heisenberg} UR, the latter must be modified to account for the detection resolution of the measurement apparatus. Another important quantum mechanical feature that one usually fails to observe from standard coarse-grained distributions is the mutual unbiasedness \cite{Durt10} relation between measurement outcomes of complementary observables. That is, eigenstates of--say--the coarse-grained position operator do not necessarily present a uniform distribution of outcomes for coarse-grained momentum measurements. Interestingly, it was shown in Ref. \cite{tasca18a} that one can indeed enjoy full quantum mechanical unbiasedness using a \textit{periodic} coarse-graining design rather than the standard one.
Other practical issues regarding false positives in entanglement detection \cite{Tasca13,Ray13a} and cryptographic security \cite{Ray13a,Ray13b} must also be reconsidered when one deals with realistic coarse-grained distributions. 

In this section, we will introduce the projective measurement operators both for the \textit{standard} and the \textit{periodic} models of coarse graining. Practical features such as measurement resolution, detector range and positioning degrees of freedom in the measurement design will be discussed. We will also briefly discuss relations of mutual unbiasedness  between coarse-grained measurement outcomes in complementary domains. A detailed discussion of uncertainty relations for coarse-grained distributions will be presented in the next section.

\subsection{Coarse-graining models}
\label{sec:CGmodels}
A laboratory experiment necessarily yields a discrete, finite set of measurement outcomes
of any observable in any physical system. This is also the case for an experiment probing a continuous degree of freedom, 
$\hat u$, for which measurement outcomes $\{u_k\}$ labeled by the discrete integer index $k\in\mathbb{Z}$ relate to the underlying continuous real variable $u\in \mathbb{R}$ corresponding to the eigenspectra of $\hat u$. In the most general scenario, a coarse-graining model is obtained from an arbitrary partition of the set of real numbers $\mathbb{R}$, in intervals $\mathcal{R}_k$ with $u_k\in \mathcal{R}_k$.
The orthogonality of the measurement outcomes requires the subsets to be mutually disjoint: $\mathcal{R}_k \cap\mathcal{R}_{k'}= \emptyset$, $\forall$  $k \neq k'$.
Even though the continuous variable can be formally discretised into an infinite number of outcomes (with $k$ an unbounded integer), the experiment can only probe a finite range of the continuous variable. Thus, the detection range, $\mathcal{R}_{\rm range}$, can be formally defined by the union of the disjoint subsets associated with the probed outcomes:
 \beq
 \label{UnionRk}
 \underset{k}{\cup}\, \mathcal{R}_k =\mathcal{R}_{\rm range} \subset \mathbb{R}.
 \eeq
This relation limits the set of possible values of $k$ to a finite subset of integers $\mathcal Z_k \subset \mathbb Z$.
Due to the finite range, $\mathcal{R}_{\rm range}$, of the measurement process
it is important to secure under reasonable experimental conditions that the underlying probability density is supported within the chosen range of detection \cite{Ray13a,Ray13b}. Mathematically, a faithful coarse-grained measurement design should ensure that
\begin{equation}\label{RangeAssumption}
\int_{\mathcal{R}_{\rm range}}  P_u(u)  du \approx 1,
\end{equation}
where $P_u$ is the marginal pdf defined in 
Eq.(\ref{marginalPDuv}).
\par
The probability $\rom p_k^{(u)}$ that the outcome $u_k$ is produced writes as an integral of the marginal probability density, $P_u$, for the continuous variable:
\begin{equation}\label{P_k}
\rom p_k^{(u)}= \int_{\mathcal{R}_k}  P_u(u)  du,
\end{equation}
where the integration is performed in the interval $\mathcal{R}_k$. 
Due to the faithful coarse-grained condition in Eq.(\ref{RangeAssumption}) we have
\beq
\label{sum-prob-approx-1}
\sum_{k\in {\cal Z}_k} \rom p_k^{(u)}\approx 1.
\eeq
We can define projective  operators associated with the coarse-grained measurements:
\begin{equation}\label{ProjectiveOperatorGeneral}
\hat{C}^{(u)}_k = \int_{\mathcal{R}_k}  |u\rangle \langle u| du, 
\end{equation}
so that the probabilities \eqref{P_k} can be written as 
\begin{equation}\label{P_kOperator}
\rom p^{(u)}_k= {\rm Tr}  (\hat{\rho}\hat{C}^{(u)}_k),
\end{equation}
with $P_u(u)=\langle u | \hat{\rho} |u \rangle$.
In order to study mutual unbiasedness and uncertainty relations, we shall later in this and the following sections define coarse-grained operators like those in Eq. \eqref{ProjectiveOperatorGeneral} for conjugate variables of the quantum state, such as the position and the linear momentum of a quantum particle.

\subsubsection{Standard Coarse Graining}

\begin{figure}[b]  
\centering
\includegraphics[width=15cm]{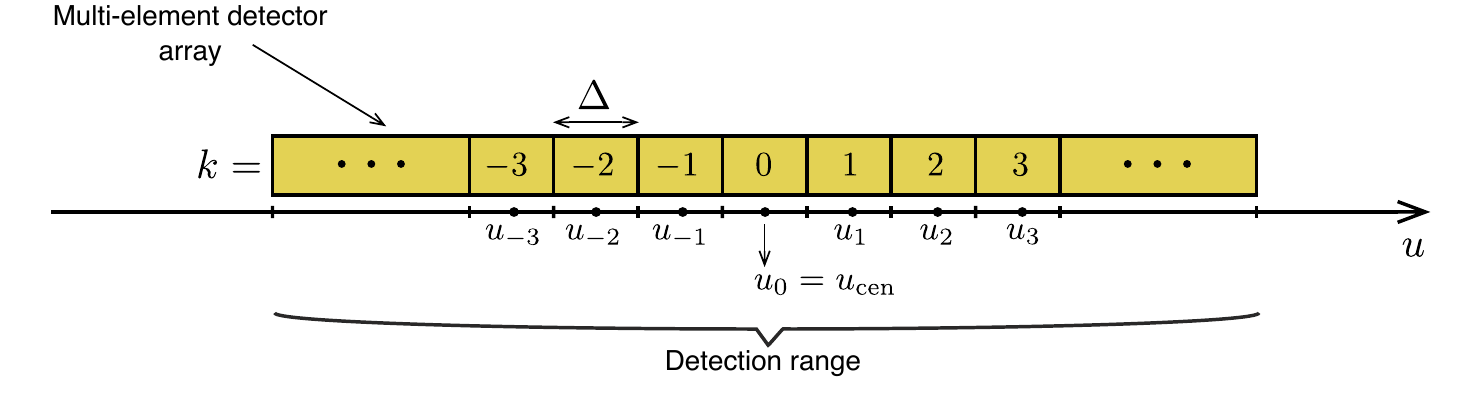} 
\caption{Multi-element detector array illustrating the standard coarse-graining geometry.}
\label{Fig:DetectorArray}
\end{figure} 

\begin{figure}[t]
\centering
\includegraphics[width=15cm]{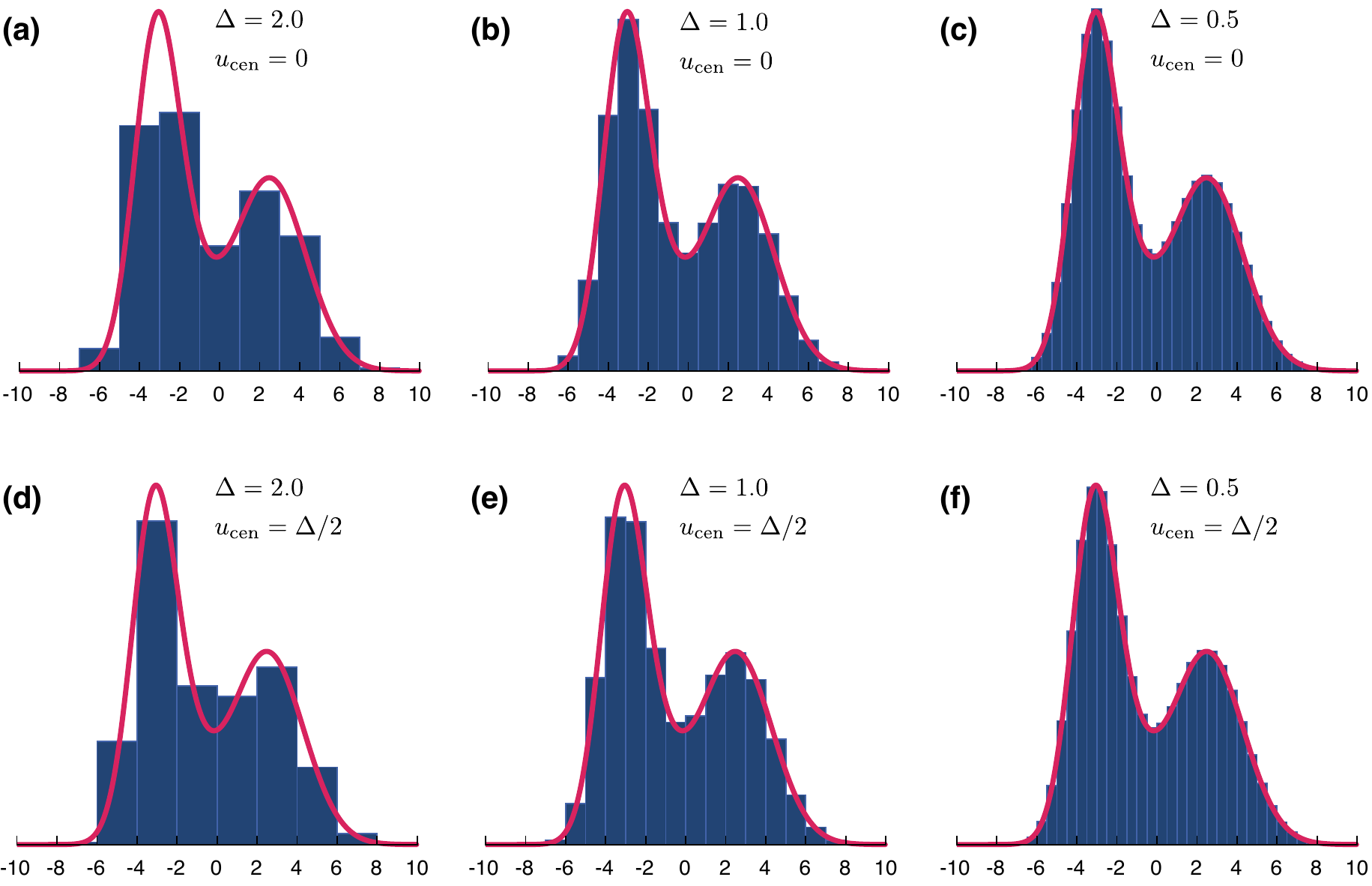}
\caption{Coarse-grained distributions (blue bars) according to the standard model. The red solid line indicates the underlying continuous distribution used to generate the discretised versions. The used resolution $\Delta$ and positioning degree of freedom $u_{\rm cen}$ is indicated beside each distribution. For each resolution, two distinct distributions are shown, each of which associated with a different positioning of the coarse-graining bins.}
\label{Fig:CGdistributions}
\end{figure}

The standard model of coarse graining describes, for example, the typical projective measurements performed with an array of adjacent, rectangular detectors. A  conventional example of such an apparatus is the image sensor of a digital camera, for which the pixel size stands for the detection resolution whereas the length of the full sensor embodies the range of detection. In the current analysis, we shall consider a linear detector array along a single spatial dimension rather than the two-dimensional area of a typical  image sensor, as illustrated in Fig. \ref{Fig:DetectorArray}. The coarse-graining interval representing the detection window of the $k$-th pixel of the linear array is then:
\beq
\label{RectangleFunction}
\mathcal{R}_k:=\left( u_{\rm cen} + (k-\frac{1}{2}) \Delta,  u_{\rm cen} + (k+\frac{1}{2})\Delta \right] ,
\eeq
where $\Delta$ is the detector or pixel size -- also commonly referred to as the \textit{coarse-graining width} or the \textit{bin width}. Using the definition \eqref{RectangleFunction}, the discretised outcomes $u_k$ represent the $u$ value of the center of the corresponding bin: 
\beq 
u_k=u_{\rm cen}+k\Delta. 
\label{eq:uk}
\eeq
The parameter $u_{\rm cen}$ sets the position of the central bin of the array, whose outcome label is $k=0$, yielding $u_0=u_{\rm cen}$. To illustrate the effect of the coarse-graining design on measured distributions, we plot in Fig. \ref{Fig:CGdistributions} coarse-grained distributions (blue bars) obtained using 3 different resolutions: $\Delta=2$ (left colum), $\Delta=1$ (central column) and $\Delta=1/2$ (right column). For each resolution, we plot two distinct distributions obtained using $u_{\rm cen}=0$  (top row) and $u_{\rm cen}=\Delta/2$  (bottom row). In other words, the coarse-graining bins of the distributions plotted at the bottom part of the figure are displaced by half a ``pixel" in relation to the distributions at the top. Clearly, the distribution obtained using a fixed resolution is not unique, but the effect of small displacements (smaller than the bin width) gets less important as the resolution is increased. For comparison, the generating continuous distribution is plotted in red.

We shall now use this model for standard coarse graining to explicitly define the discretised counterparts of the position and momentum operators given in Eq. \eqref{defuv}. 
\begin{subequations}\label{PosMomOperators}
\begin{align}
\hat{u}_\Delta&= \sum_k u_k \hat C_k^{(u)} ,  		\label{PosOperator}\\
\hat{v}_\delta&= \sum_l v_l \hat C_l^{(v)}, 		 \label{MomOperator}
\end{align}
\end{subequations}
where the projector $\hat C_k$ is defined in Eq.(\ref{ProjectiveOperatorGeneral}) (with $\hat C_l^{(v)}$ having an equivalent definition for $\hat v$ measurements), and we used $\Delta$ ($\delta$) as the detection resolution for $\hat u$ ($\hat v$) measurements. According to the definition in Eq.\eqref{P_k}, as a result of the the coarse-grained measurement of $\hat u$ and $\hat v$ we obtain the discrete probabilities, $\rom p^{(u)}_{\Delta, k}$ and  
$\rom p^{(v)}_{\delta,l}$.
The discrete variances associate with these discrete probabilities are:
\begin{subequations}\label{DiscreteVar}
\begin{align}
\sigma_{P_\Delta^{(u)}}^2 &= \sum_k u_k^2\; \rom p_{\Delta,k}^{(u)} -\left(  \sum_k u_k\; \rom p_{\Delta,k}^{(u)}\right)^2,  		\label{DiscreteVarPos}\\
\sigma_{P_\delta^{(v)}}^2 &= \sum_l   v_l^2 \;\rom p_{\delta ,l}^{(v)} -\left(  \sum_l v_l 
\;\rom p_{\delta,l}^{(v)}\right)^2,	 \label{DiscreteVarPos}
\end{align}
\end{subequations}
where we define the set of discrete probabilities:
\beq
P_{\Delta}^{(u)}:=\{\rom p_{\Delta,k}^{(u)}\}\;\;\;\mbox{and}\;\;\;
P_\delta^{(v)}:=\{\rom p_{\delta,k}^{(v)}\}.
\eeq
One can see from the definitions \eqref{DiscreteVar} that if the bin widths $\Delta$ and $\delta$ are such that $\rom p_{\Delta,k}^{(u)} $ and $\rom p_{\delta,l}^{(v)} $ are sufficiently close to unity for for some value of $k$ and $l$, we have 
$\sigma_{P_\Delta^{(u)}}^2,\sigma_{P_\delta^{(v)}}^2\longrightarrow 0$.  Thus, naive application of any of the variance-based URs given in section \ref{sec:varianceURs} would indicate a false violation of a UR.  It has been shown in Ref. \cite{rudnicki12a} that the same argument applies to discretized versions of entropic URs, such as those of section \ref{EntropicUR}.  Thus, proper treatment of standard coarse-grained measurements is essential in order to take advantage of the practical application of URs in QIT and quantum physics in general.  In section \ref{sec:CGUR} we show how this can be done.

\subsubsection{Periodic Coarse Graining}

\label{PCGsect}
\begin{figure}[b]  
\centering
\includegraphics[width=15cm]{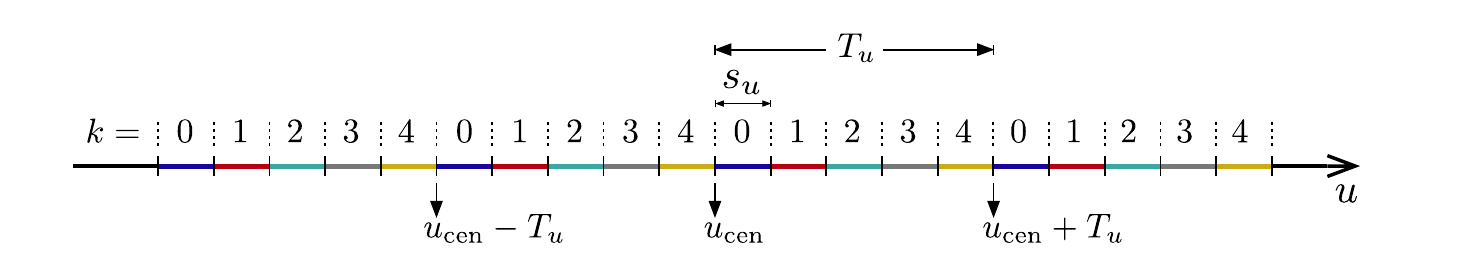} 
\caption{Periodic coarse-graining design with $d=T_u/s_u=5$ detection outcomes. The parameter $T_u$ is the periodicity in which bins of size $s_u$ are arranged.}
\label{Fig:PeriodicCGScheme}
\end{figure} 

A distinct model of coarse graining discussed in the literature \cite{tasca18a,paul18a} is refereed to as \textit{periodic coarse graining} (PCG). In this model, the partition of the whole set of real numbers $\mathbb{R}$ is performed in a periodic manner, leading to a finite number $d$ of subsets $\mathcal{R}_k$, with $k=0, \cdots, d-1$. The resulting discretization utilizes the index $k$ as a direct label for the detection outcomes, in a similar fashion to what is usually defined for finite-dimensional quantum systems. The subsets  $\mathcal{R}_k$ are defined as \cite{tasca18a}:
\beq
\label{PeriodicFunction}
\mathcal{R}_k:=\left\{ u \in \mathbb{R} \, | \,  u_{\rm cen}+ ks_u \leqslant u ({\rm mod} \, T_u) <  \ u_{\rm cen} + (k+1)s_u \right\},
\eeq
where $s_u$ plays the role of a bin width similar to the resolution $\Delta$ used for the standard coarse graining. In the definition \eqref{PeriodicFunction}, bins of size $s_u$ are arranged periodically with the parameter $T_u$ representing the period, as illustrated in Fig. \ref{Fig:PeriodicCGScheme} for the particular design using $d=T_u/s_u=5$ detection outcomes. It is important to notice that this coarse graining design do not distinguish detections in distinct bins associated with the same detection outcome $k$ (ranging from $0$ to $4$ in Fig. \ref{Fig:PeriodicCGScheme}). For example, a detection within any bin colored in red in Fig. \ref{PeriodicFunction} would lead to the same detection outcome $k=1$.

An interesting feature of the PCG model is that the number of detection outcomes is utterly adjustable by the choice of the parameters $T_u$ and $s_u$, regardless of the chosen detection range. For instance, doubling the range of detection allows one to design PCG measurement using twice as much periods in its design, while maintaining the same number $d=T_u/s_u$ of detection outcomes. 
As with the standard model, the reference coordinate $u_{\rm cen}$ sets the center of the detection range also for the PCG design. Using the subset definition given in Eq. \eqref{PeriodicFunction}, we can explicitly write the projector operators, Eq. \eqref{ProjectiveOperatorGeneral}, for the PCG model as
\begin{equation}\label{ProjectiveOperatorPeriodicU}
\hat{\Pi}^{(u)}_k = \int_{\mathcal{R}_k}  |u\rangle \langle u| du = \sum_{n \in \mathbb{Z}}  \int_{u_{\rm cen} + ks_u + nT_u}^{u_{\rm cen} + (k+1)s_u + nT_u}  |u\rangle \langle u| du,
\end{equation}
where we extend the sum in $n$ over $\mathbb{Z}$ without loss of generality, assuming that Eq. \eqref{RangeAssumption} is satisfied. Analogously, we also define the PCG projective operators over the conjugate variable $v$:
\begin{equation}\label{ProjectiveOperatorPeriodicV}
\hat{\Pi}^{(v)}_l = \int_{\mathcal{R}_l}  |v\rangle \langle v| dv = \sum_{n \in \mathbb{Z}}  \int_{v_{\rm cen} + ls_v + nT_v}^{v_{\rm cen} + (l+1)s_v + nT_v}  |v\rangle \langle v| dv,
\end{equation}
where we define $s_v$ and $T_v$ as the bin width and periodicity used in the PCG measurements of $v$.

\subsection{Mutual unbiasedness in coarse-grained measurements}
\label{MUBsect}
If a quantum system is described as an eigenstate of a given observable, the measurement outcomes of a complementary observable are completely unbiased: each one of them occurring with equal probability, $1/d$, where $d$ is the dimension of the quantum system's Hilbert space. This unbiasedness relation is an important feature of quantum mechanics with no classical counterpart, and is usually cast in terms of the basis vectors constituting the eigenstates of two (or more) complementary observables.  To be more precise, two orthonormal bases $\{ |a_k\rangle \}$ and $\{ |b_l\rangle \}$ are said to be \textit{mutually unbiased} if and only if $| \langle a_k | b_l \rangle|^2 =1/d$ for all $k,l=0,\cdots,d-1$ \cite{Durt10}. 
The observation of unbiased measurement outcomes is customary in experiments with finite dimensional quantum systems. Not only routine, measurements in mutually unbiased bases (MUB) constitute a key procedure in a number of quantum information processing tasks, such as verification of cryptographic security \cite{Coles2017}, certification of quantum randomness \cite{vallone14}, detection of quantum correlations \cite{spengler12,krenn14,erker17} and tomographic reconstruction of quantum states \cite{Fernandez-Perez11,giovannini13}.

Mutual unbiasedness is also extendable to continuous variables quantum systems \cite{Weigert08}, for which conjugate bases $\{ |u\rangle \}$ and $\{ |v\rangle \}$ satisfy  $| \langle u | v \rangle|^2 =1/2\pi\hbar$, \textit{i.e.}, the overlap of the basis vectors $|u\rangle$ and $|v\rangle$ is independent  (no bias) of their eigenvalues, $u$ and $v$. For CV systems, nevertheless, this relation is rather a theoretical definition than an experimentally observable fact, since the experimentalist has neither the capability to prepare nor to measure the (infinitely squeezed) eigenstates of the $\hat{u}$ and $\hat{v}$. Instead, both the preparation and measurement procedures are limited to the finite resolution of the experimental apparatus. As discussed previously in this section, measurements of a CV degree of freedom render discretized, coarse-grained outcomes whose probabilities, Eq. \eqref{P_k}, are provided by a coarse-graining model described by the projective operators given in Eq. \eqref{ProjectiveOperatorGeneral}. These coarse-grained probabilities obtained experimentally do not in general preserve the mutual unbiasedness complied by the underlying continuous variables. 

To elaborate the issue, let us consider sets of projectors $\{ \hat{C}^{(u)}_k \}$ and $\{ \hat{C}^{(v)}_l \}$ defining coarse-graining measurements in the complementary domains $u$ and $v$ of a continuous variable quantum system $\hat{\rho}$. We assume measurement designs providing a number $d$ of outcomes in each domain. In this scenario, the requirement for mutual unbiasedness is thus that the coarse-grained probabilities for measurements of one variable are evenly spread between all discretized outcomes whenever the quantum state is localized with respect to the coarse graining applied to its conjugate variable (and vice-versa). The subtlety in this requirement is the (infinite) degeneracy of normalizable quantum states that can be localized with respect to the chosen coarse graining. To emphasize this degeneracy, we refer to the outcome probabilities, Eq. $\eqref{P_k}$, with explicit dependency on the quantum state in order to mathematically phrase the condition for mutual unbiasedness in coarse-grained CV:  the outcomes of $\{ \hat{C}^{(u)}_k \}$ and $\{ \hat{C}^{(v)}_l \}$ are mutually unbiased if \textit{for all} quantum states $\hat{\rho}$ and $k_0,l_0=0,\cdots, d-1$ we have \cite{tasca18a}: 
\begin{subequations} \label{MUB_CG}
\beq \label{MUB_CG_pos}
\rom p^{(u)}_k (\hat{\rho}) =\delta_{k_0k} \hspace{0.1in} \Rightarrow   \hspace{0.1in}  \rom p^{(v)}_l (\hat{\rho}) =d^{-1},
\eeq
\beq \label{MUB_CG_mom}
\rom p^{(v)}_l (\hat{\rho}) =\delta_{l_0l}   \hspace{0.1in}  \Rightarrow   \hspace{0.1in}   \rom p^{(u)}_k (\hat{\rho}) =d^{-1},
\eeq
\end{subequations}
where, again, we stress that $\rom p^{(u)}_k (\hat{\rho}) ={\rm Tr}  (\hat{\rho}\hat{C}^{(u)}_k)$ and $\rom p^{(v)}_l (\hat{\rho}) ={\rm Tr}  (\hat{\rho}\hat{C}^{(v)}_l)$, as in Eq. \eqref{P_k}.

Having formulated the conditions for mutual unbiasedness, Eqs. \eqref{MUB_CG}, it is easy to perceive that the adjacent, rectangular subsets defining the standard coarse graining [Eq. \eqref{RectangleFunction}] will not lead to unbiased measurement outcomes. Any CV distribution localized in a single coarse-graining bin (for example in the $u$ variable) generates a probability density that decays in the Fourier domain (the $v$ variable) along the adjacent bins within the detection range. This decay generates a non constante coarse-grained distribution that, by definition, is biased. Furthermore, the number $d$ of detection outcomes in the standard design depends directly on the selected detection range, as well as on the chosen resolution. As a consequence, even though a particular localized distribution could lead to approximately unbiased coarse-grained outcomes in the Fourier domain, an extended detection range would increase the number of outcomes, thus spoiling the unbiasedness. 

It is thus evident that in order to retrieve unbiased outcomes from coarse-grained measurement, a more contrived coarse-graining design is needed. As it turns out, it was shown in Ref. \cite{tasca18a} that the PCG design exactly fulfils the requirements for unbiased measurements of finite cardinality stated in Eqs \eqref{MUB_CG}. A relation between the periodicities $T_u$ and $T_v$ used in the PCG of the conjugate variables $u$ and $v$ was analytically derived as a single condition for unbiased coarse-grained measurements:
\beq \label{MUBcondition}
\frac{T_uT_v}{2\pi \hbar}= \frac{d}{m} , \hspace{0.2in } m \in \mathbb{N} \hspace{0.2in} \rom s.t.  \hspace{0.2in} \forall_{n=1, \cdots, d-1}  \hspace{0.1in}  \frac{mn}{d} \notin \mathbb{N}.
\eeq
The unbiasedness condition stated in Eq. \eqref{MUBcondition} establishes infinite possibilities for the pair of periodicities $T_u$ and $T_v$ that can be used to design the mutually unbiased pair of PCG measurements defined in Eqs. \eqref{ProjectiveOperatorPeriodicU} and \eqref{ProjectiveOperatorPeriodicV}, respectively. For instance, the simplest and most important case is the condition with $m=1$, since it is valid for all $d$ and provides the best trade-off between experimentally accessible periodicities: $T_uT_v=(2\pi \hbar) d$. Conditions with $m>1$ are also possible but are not general since they depend on the chosen number of outcomes $d$ \cite{tasca18a}. For example, for $d=4$, valid conditions are found using $m ( \textrm{mod} \, d)=1,3$ whereas for $d=5$, valid conditions are found using $m ( \textrm{mod} \, d)=1,2,3,4$. Importantly, the case with $m ( \textrm{mod} \, d)=0$ is always excluded, since in this case the PCG projectors describe commuting sets, $\left[\hat{\Pi}^{(u)}_k,\hat{\Pi}^{(v)}_l \right]=0$,  $\forall$ $k,l$ \cite{aharanov69,busch86,reiter89}.
In other words a joint eigenstate of the product $\hat{\Pi}^{(u)}_k\hat{\Pi}^{(v)}_l$  existis for all $k$ and $l$ whenever $T_uT_v=2\pi \hbar/c$ with $c \in \mathbb{N}$ \cite{rudnicki16}.
It is also interesting to note that using the periodicity definition from the PCG design ($T=ds$), it is possible to write the unbiasedness condition given in Eq. \eqref{MUBcondition} in alternative, equivalent ways:
\beq \label{MUBconditionAlternative}
(a) \, T_uT_v = \frac{2\pi \hbar}{m} d, \hspace{0.2in } (b) \,  T_us_v =\frac{2\pi \hbar}{m}, \hspace{0.2in} (c) \,  s_uT_v =\frac{2\pi \hbar}{m}, \hspace{0.2in} (d) \,  s_us_v =\frac{2\pi \hbar}{m} \frac{1}{d}.
\eeq

Finally, in Ref. \cite{paul18a} these results were generalized for PCG measurements applied to an arbitrary pair of phase space variables other than the conjugate pair formed by position and momentum. What is more, a triple of unbiased PCG measurements was also shown to exist for rotated phase space variables, along the same lines as the demonstration of a MUB triple in the continuous regime done in Ref. \cite{Weigert08}. Experimental demonstrations of unbiased PCG measurements were also carried out in Refs. \cite{tasca18a,paul18a}, both of them utilizing the transverse spatial variables of a paraxial light field.

\section{UR for coarse-grained observables}
\label{sec:CGUR}
A kind of a paradigm shift in the theory of
uncertainty relations was brought by the observation that everything
can be efficiently characterized solely by means of probability distributions.
As a result, tools known from information theory, such as information
entropy, Fisher information and other measures, came into play. Additionally,
the notion of uncertainty for discrete systems could better be captured
that way. Since products of variances calculated for
observables such as the spin are bounded in a state-dependent manner
(so that the ultimate lower bound typically assumes the trivial value of $0$), information entropies
provide an attractive alternative \cite{deutsch83}. Written already
in the R{\'e}nyi form, 

\begin{equation}
H_{\alpha}\left[P\right]=\frac{1}{1-\alpha}\ln\sum_{k}[\rom p_{k}]^{\alpha},
\end{equation}
the above equation is a discrete counterpart of Eq. (\ref{RenyiEntropy}), which corresponds to the discrete counterpart of Eq.\eqref{defShannonEntropy} when $\alpha=1$. 
\par
In the finite-dimensional case given by an arbitrary state $\hat \rho$ acting on 
a $d$-dimensional Hilbert space $\mathcal{H}$, and a pair of
non-degenerate, non-commuting observables, $\hat A$ and $\hat B$, one usually
defines the probabilities:
\begin{equation}
\rom p_{i}^{(A)}=\left\langle a_{i}\right|\hat \rho\left|a_{i}\right\rangle ,\qquad \rom p_{j}^{(B)}=\left\langle b_{j}\right|\hat \rho\left|b_{j}\right\rangle .\label{PrDys}
\end{equation}
By $\left|a_{i}\right\rangle $ and $\left|b_{j}\right\rangle $,
$i,j=1,\ldots,d$ we denote the eigenstates of the operators associated
with both observables. Disctrete entropic URs for the above probability
distributions are of the general form 

\begin{equation}
H_{\alpha}\left[P^{(A)}\right]+H_{\beta}\left[P^{(B)}\right]\geq B_{\alpha\beta}\left(\mathbf{U}\right),\label{De}
\end{equation}
with $\mathbf{U}\in\mathcal{U}\left(d\right)$ being a unitary matrix with
matrix elements $\mathbf{U}_{ij}=\left\langle a_{i}\left|b_{j}\right\rangle \right.\!$.
We denote $P^{(A)}:=\{\rom p_i^{(A)} \}$ and $P^{(B)}:=\{\rom p_j^{(B)} \}$ again with $i,j=1,\ldots,d$.

The first entropic uncertainty relation for discrete variables comes from Deutsch
\cite{deutsch83}, who for $\alpha=1=\beta$ found the lower bound $B_{11}^{\textrm{D}}=-2\ln C$,
with $C=\left(1+\sqrt{c_{1}}\right)/2$ and $c_{1}=\max_{i,j}\left|U_{ij}\right|^{2}$.
A substantially more renowned Maassen\textendash Uffink (MU) bound \cite{maassen88}
derived in 1988, is $B_{\alpha\beta}^{\textrm{MU}}=-\ln c_{1}$. This
bound is however valid only for the conjugate parameters $1/\alpha+1/\beta=2$.
Very recently, a plethora of new results \cite{korzekwa2014,friedland2013,puchala2013,coles2014,RPZ2014,bosyk2014a,bosyk2014b,kaniewski2014,PRKZ2018}
improving the celebrated MU bound has been obtained. In particular,
an approach based on the notion of majorization (suitable from the
perspective of resource theories and quantum thermodynamics \cite{Brandao3275})
provides a significant qualitative novelty \cite{friedland2013,puchala2013,RPZ2014,PRKZ2018}, which
will also be touched upon in this section.

In this review we are concerned with the case in which continuous
probability distributions $P_{u}\left(u\right)$ and $P_{v}\left(v\right)$
are replaced (\textit{viz.} they were measured this way) by their discrete counterparts
($k,l\in\mathbb{Z}$). According to the discussion in Section \ref{Section:realistic-CG-models} we can use the definitions in Eq.(\ref{P_k}) and 
\eqref{RectangleFunction}, and the condition in Eq.\eqref{UnionRk}, to write the discrete probabilities: 
\begin{equation}
\rom p_{\Delta,k}^{(u)}=\int_{\left(k-1/2\right)\Delta}^{\left(k+1/2\right)\Delta}dy\;P_{u}\left(y\right),\qquad 
\rom p_{\delta,l}^{(v)}=\int_{\left(l-1/2\right)\delta}^{\left(l+1/2\right)\delta}dy\;P_{v}\left(y\right),
\label{dyskretne2}
\end{equation}
with  $k \in \mathcal Z_k \subset \mathbb Z$. In the following we describe a series of URs 
for these discrete probabilities that are known as coarse-grained URs, derived in \cite{bialynicki84,bialynicki06,rudnicki12a,rudnicki12b}. These are the coarse-grained counterpart of the {\it Heisenberg}, {\it Shannon entropy} and {\it R\'enyi entropy} URs
in Eqs.(\ref{VarianceUR}),(\ref{ShannonUR}) and (\ref{RenyiUR}) respectively.
Here, we will closely  follow the treatment  in  \cite{rudnicki12a,rudnicki12b}, however, before we start we give a short historical overview and discuss a path towards extensions going beyond CCOs. 

The idea that generic quantum uncertainty could be quantified by the sum of Shannon entropies evaluated for discretized position and momentum probability distributions for the first time appeared in the contribution by Partovi \cite{partovi83}. He also derived the first coarse-grained UR which in the form is reminiscent\footnote{Note that both papers \cite{deutsch83,partovi83} have been published in 1983, however, Partovi  in his first sentence refers to a ''recent letter'' by Deutsch.} to the Deutsch bound for finite-dimensional systems \cite{deutsch83}. Both bounds \cite{deutsch83,partovi83} were obtained by means of a direct optimization, independently applied to every logarithmic contribution. Symmetry in developments of the URs for finite-dimensional and coarse-grained systems happened to be much deeper as the second coarse-grained result, by Bialynicki-Birula \cite{bialynicki84}, is a counterpart of the MU bound \cite{maassen88}. The former result is an application of the continuous variant of the {\it Shannon entropy} UR (so the $L_p$-$L_q$ norm inequality by Beckner \cite{beckner1975}) supported by the Jensen inequality for convex functions, while the MU bound is a direct consequence of the Riesz theorem for the $l_p$-$l_q$ norms. Note that relatively often, integration limits in \eqref{dyskretne2} were chosen as ''from $k \Delta$ to $(k+1)\Delta$'' and ''from $l \delta$ to $(l+1)\delta$'', however this choice causes a formal pathology in the limit of infinite coarse graining \cite{Rudnicki2011JLR}. Thus, sticking to terminology of Eq. \eqref{RectangleFunction}, in theory it is better to avoid borderline settings for the position of the central bin, i.e. $u_{\textrm{cen}}=\Delta/2$.

To briefly report later developments, one shall mention that Partovi reconsidered the problem he had posed several years ago, pioneering applications of majorizaiton techniques \cite{partovi2011}. Also Sch{\"u}rmann  and Hoffmann \cite{Schurmann2009} discussed the {\it Shannon entropy} UR from the perspective of the integral equation associated to it, while the first author conjectured an improvement (later mentioned in detail) which agrees with his numerical tests \cite{Schurmann2012}. Finally, we mention (without details) an erroneous improvement of \cite{bialynicki06} by Wilk and Wlodarczyk \cite{wilk2009,IBB2010}, mainly devoted to the case of the Tsallis entropy.

Although originally the URs were derived for CCOs, $\hat u$ and $\hat v$, here we show 
which of the URs in \cite{rudnicki12a,rudnicki12b} can be valid also 
for operators $\hat u$ and $\hat v$ 
that are arbitrary linear combinations of all positions and momenta 
of the $n-$bosonic modes like the ones defined in Eq.\eqref{defuv}, {\it viz.} operators that are not necessarily CCOs.  
In the general case, we stress that there is always a unitary metaplectic transformation\footnote{So $\hat U_{\mathbf{S}}$ belongs to the metaplectic group 
$Mp(2n,\mathbb{R})$ and it is always associated with a matrix $\mathbf{S}$ that belongs 
the symplectic group $Sp(2n,\mathbb{R})$ \cite{Dutta1995}.}, $\hat U_{\mathbf{S}}$, that  
connects $\hat u$ and $\hat v$, {\it viz.} $\hat v=\hat U_{\mathbf{S}}^\dagger \hat u \hat U_{\mathbf{S}}$.
However, this metaplectic transformation is not necessarily a $\pi/2$ rotation, which would be the case if $\hat u$ and $\hat v$ were CCOs. 
In order to see this, we first define two sets of operators $({\bf \hat u},{\bf \hat u}^\prime)^T=(\hat u=\hat u_1,\ldots,\hat u_n,\hat u_1^\prime,\ldots, \hat u_n^\prime)^T=\sqrt{\gamma}\,\tilde{\mathbf{S}}\, {\bf \hat x}$ and $({\bf \hat v},{\bf \hat v}^\prime)^T=(\hat v=\hat v_1,\ldots,\hat v_n,\hat v_1^\prime\ldots, \hat v_n^\prime)^T=
\sqrt{\gamma}\,{\mathbf{S}^\prime}\, {\bf \hat x}$, where $\tilde{\mathbf{S}}$ and $\mathbf{S}^\prime$ are some matrices belonging to the symplectic group $Sp(2n,\mathbb{R})$,
with the only restriction that the first rows of $\tilde{\mathbf{S}}$ and $\mathbf{S}^\prime$ correspond to the real coefficients ${\bf d}$ and ${\bf d}^\prime$ in Eq.(\ref{defofd}), respectively, which define the operators $\hat u$ and $\hat v$ in Eq.(\ref{defuv}).
Due to the properties of symplectic matrices, all the pairs $\hat u_i$ and $\hat u_j^\prime$, and also $\hat v_i$ and $\hat v_j^\prime$, satisfy CCRs, {\it viz.} $[\hat u_i,\hat u_j^\prime]=i\hbar \gamma \delta_{ij}$ and $[\hat v_i,\hat v_j^\prime]=i\hbar \gamma\delta_{ij}$
with $i,j=1,\ldots,n$. But it is immediate to see that $({\bf \hat v},{\bf \hat v}^\prime)^T=\mathbf{S}({\bf \hat u},{\bf \hat u}^\prime)^T$
where the matrix $\mathbf{S}:={\mathbf{S}}^\prime\tilde{\mathbf{S}}^{-1}$ is a generic symplectic matrix. 
Then the Stone-von-Neumann theorem guarantees that the change $({\bf \hat u},{\bf \hat u}^\prime)^T\rightarrow ({\bf \hat v},{\bf \hat v}^\prime)^T$ is unitarily implementable
by a metaplectic transformation $\hat U_{\mathbf{S}}$ \cite{Dutta1995}. In particular we have
$\hat U_{\mathbf{S}}^\dagger \,\hat u\,\hat U_{\mathbf{S}}=(\mathbf{S}\,{\bf \hat x})_1=:\hat v$.
\par
\subsection{URs proved only for  CCOs}
\par
The key concept behind the treatment of coarse-grained URs in \cite{rudnicki12a,rudnicki12b}
is the introduction of the piece-wise continuous probability density functions:
\beq
Q_{\Delta,u}(u):=\sum_{k\in \mathcal Z_k} \rom p^{(u)}_{\Delta,k}\;D_\Delta(u,u_k)\;\;\;
\label{eq:Qu}
\mbox{and}\;\;\;
Q_{\delta,v}(v):=\sum_{l \in \mathcal Z_l} \rom p^{(v)}_{\delta,l}\;D_\delta(v,v_l),
\eeq
where $D_\Delta(u,u_k)$ and $D_\delta(v,v_l)$ are called the histogram functions
(HF) with $u_k$ (and $v_l$ in an analogous way) defined in Eq. \eqref{eq:uk}. Generically, these functions are defined such that they are normalized in each bin:
\beq
\int_{(k-1/2)\Delta}^{(k+1/2)\Delta} D_\Delta(u,u_k)\;du=1\;\;\;\mbox{and}\;\;\;
\int_{(l-1/2)\Delta}^{(l+1/2)\Delta} D_\Delta(v,v_l)\;dv=1,
\eeq 
and approach the Dirac delta distribution for infinitesimal bin size:
\beq
\lim_{\Delta \rightarrow 0} D_\Delta(u,u_k)=\delta(u-u_k)\;\;\;\mbox{and}\;\;\;
\lim_{\delta \rightarrow 0} D_\delta(v,v_l)=\delta(v-v_l). \label{limitsD}
\eeq
Therefore, in the limit $\mathcal Z_k,\mathcal Z_l\rightarrow \mathbb Z$
and $\Delta,\delta\rightarrow 0$ we have 
$Q_{\Delta,u}(u)\rightarrow P_u(u)$ and $Q_{\delta,v}(v)\rightarrow P_v(v)$.
We shall stress here that the HF can, in general, have any functional form as long as it is non-negative, normalized and fulfills Eq. \eqref{limitsD}.
However, the most common histogram function is
the rectangular HF:
\beq
D^{R}_\Delta(u,u_k):= \left\{  \begin{matrix} 
      1/\Delta & \mbox{for $u\in\left((k-\frac{1}{2}) \Delta,  (k+\frac{1}{2})\Delta\right] $} \\
       0 & \mbox{otherwise.} \\
   \end{matrix}\right.,
\eeq
with an equivalent definition for $D^{R}_\delta(v,v_l)$. In Fig.\ref{Fig:CGdistributions}
we show an example of coarse-grained probability distributions functions 
$Q_{\Delta,u}(u)$ (the area beneath these functions are 
displayed in full) using rectangular histogram functions and for different size bins $\Delta$.
\par 
Here, we generalise the results in  \cite{rudnicki12a,rudnicki12b} through  
the following expression that will be justified later:
\beq
\label{generalresult}
h_\alpha[Q_{\Delta,u}]+h_\beta[Q_{\delta,v}]\geq \ln\left(\frac{\pi \hbar |\gamma|\;e^{h_\alpha[D_\Delta]-\ln\Delta+h_\beta[D_\delta]-\ln\delta}}{\varepsilon_\alpha(\Gamma/4)}\right),
\eeq
with  $1/\alpha+1/\beta=2$ and $1/2\le\alpha \le1$. To simplify the notation we define the function:
\beq\label{veps}
\varepsilon_\alpha\left(x\right):=\min\left\{\alpha^{\frac{1}{2-2\alpha}}\beta^{\frac{1}{2-2\beta}},\frac{1}{2}R^2_{00}\left(x,1\right)\right\},
\eeq
where $R_{00}(x,y)$ denotes one of the radial prolate
spheroidal wave functions of the first kind \cite{abramowitz}, and introduce the joint coarse-graining parameter $\Gamma=\Delta\delta/(\hbar|\gamma|)$.   We stress that \eqref{generalresult} involves the differential R\'enyi entropies of the piece-wise continuous distributions defined in Eqs. \eqref{eq:Qu}.
\par
\begin{figure}[t]
\centering
\includegraphics[width=7.5cm]{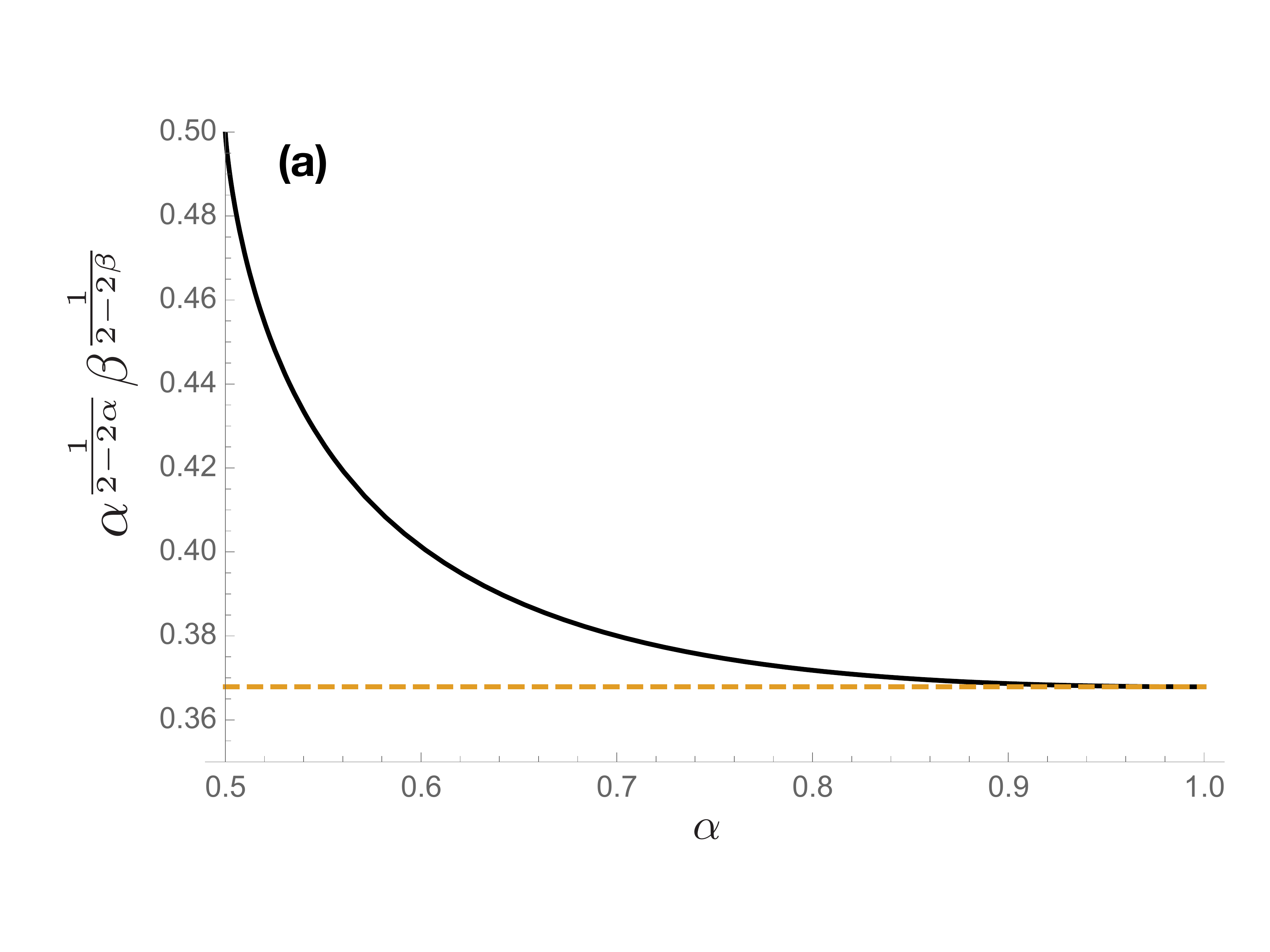}
\includegraphics[width=7.5cm]{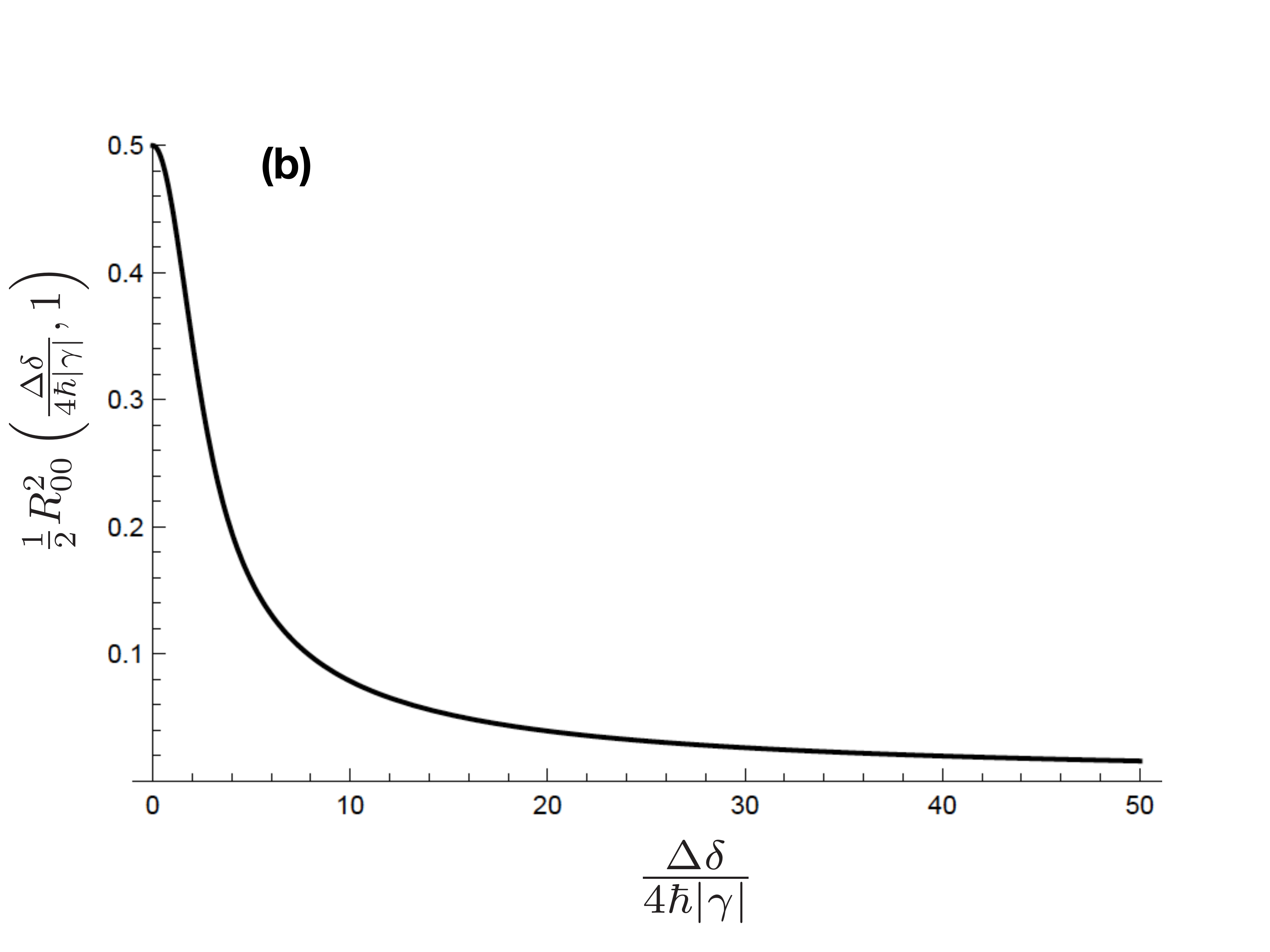}
\caption{In panel {\bf (a)} the full line is the graph of the function $f(\alpha)=\alpha^{\frac{1}{2-2\alpha}}\beta^{\frac{1}{2-2\beta}}$, with $0<\alpha\le 1$, and where $\beta(\alpha)=\alpha/(2\alpha-1)$ that stems from the condition $1/\alpha+1/\beta =2$.
The horizontal dashed line is drawn to indicate the limit $\lim_{\alpha \rightarrow 1} f(\alpha)=1/e$. In panel {\bf (b)} we plot 
the behaviour of $g(y)=(1/2) R_{00}(y,1)$ as a function of $y:=\Delta\delta/(4\hbar|\gamma|)$. 
Although $g(y)$ is shown in the range $0\le y\le 50$, it is important to note that $g(y)$ is continuous monotonically decreasing function in the positive real axis such 
$\lim_{y\rightarrow \infty} g(y)=0$.}
\label{Fig3}
\end{figure}

Let us see how the results in \cite{bialynicki84,bialynicki06,bialynicki11,rudnicki12a,rudnicki12b} can be derived from 
Eq.\eqref{generalresult}. First, we observe that 
the R\'enyi entropies of rectangular HFs, for every values of $\alpha$ and $\beta$, are:
\beq
\label{entropies-rec-HF}
h_\alpha[D^{R}_\Delta]=\ln\Delta\;\;\;\mbox{and}\;\;\;
h_\beta[D^{R}_\Delta]=\ln\delta,
\eeq
so Eq.(\ref{generalresult}) reduces to:
\beq
\label{generalresultRF}
h_\alpha[Q_{\Delta,u}]+h_\beta[Q_{\delta,v}]\geq \ln\left(\frac{\pi \hbar|\gamma| }{\varepsilon_\alpha(\Gamma/4)}\right).
\eeq
If we perform the limit $\Gamma/4 \rightarrow 0$ in Eq.(\ref{generalresultRF}), we have 
$(1/2)R^2_{00}\left(\Gamma/4,1\right)\rightarrow 1/2$, and considering 
that $1/e<\alpha^{\frac{1}{2-2\alpha}}\beta^{\frac{1}{2-2\beta}}\le 1/2$ when $1/2<\alpha\le 1$ (see Fig.(\ref{Fig3})) we recover the {\it R\'enyi-entropy} UR in Eq.(\ref{RenyiUR})
and when $\alpha=1$ the  {\it Shannon} UR in Eq.(\ref{ShannonUR}).
\par
Now, we can decompose the differential R\'enyi entropies in the left hand side of 
Eq.(\ref{generalresult}) as (see Appendix \ref{AppendixA}): 
\beq
\label{decomdiffRenyi}
h_\alpha[Q_{\Delta,u}]=H_\alpha\left[P_\Delta^{(u)}\right]+h_{\alpha}\left[D_\Delta\right]
\;\;\;\mbox{and}\;\;\;
h_\beta[Q_{\delta,v}]=H_\beta\left[P_\delta^{(v)}\right]+h_{\beta}\left[D_\delta\right],
\eeq
where we denote the set of discrete probabilities appearing in Eq.\eqref{dyskretne2} as
$P_\Delta^{(u)}:=\{\rom p_{\Delta,k}^{(u)}\}$ and $P_\delta^{(v)}:=\{\rom p_{\delta,k}^{(v)}\}$, respectively. Note that, for pdfs with bounded support, the R\'enyi entropy is maximized for the uniform distribution \cite{Lassance2017}, so we always have: $h_{\alpha}\left[D_\Delta\right]\leq \ln(\Delta)$
and $h_{\beta}\left[D_\delta\right]\leq \ln(\delta)$.
If we apply the result Eq.\eqref{decomdiffRenyi} to the inequality \eqref{generalresult} we recover the result proved in Ref.  \cite{rudnicki12b} for the discrete entropies:
\beq
H_\alpha[P_{\Delta}^{(u)}]+H_\beta[P_{\delta}^{(v)}]\geq 
\ln\left(\frac{\pi }{\varepsilon_\alpha(\Gamma/4)\Gamma}\right).
\label{CGRUR}
\eeq
This is the coarse-grained version of the {\it R\'enyi entropy} UR\footnote{Sch{\"u}rmann conjectured \cite{Schurmann2012}  that $\varepsilon_1\left(z\right)$ defined in \eqref{veps}, in the context of Eq. \eqref{CGRUR} could be replaced by  $e^{-1}R^2_{00}\left(2 z/e,1\right)$.} in Eq.\eqref{RenyiUR}.
We shall also emphasize, as the title of this subsection suggests, that the demonstration of the URs \eqref{CGRUR} presented in Ref. \cite{rudnicki12b} uses explicitly the fact that $\hat u$ and $\hat v$ form a CCO pair.
Therefore, the UR in Eq.\eqref{generalresult} is, in principle,  valid only for 
CCO pairs, since it can be obtained from Eq.\eqref{CGRUR}  by adding $h_{\alpha}\left[D_\Delta\right]+h_{\beta}\left[D_\delta\right]$ to both sides, and using  Eq.(\ref{decomdiffRenyi}).
\par
The discrete R\'enyi entropy is always positive, and we have
\beq
\lim_{\Gamma\rightarrow +\infty} 
\ln\left(\frac{\pi }{\Gamma \varepsilon_\alpha(\Gamma/4)}\right)=
\lim_{\Gamma\rightarrow +\infty} 
\ln\left(\frac{\pi }{
\frac{1}{2} \Gamma R^2_{00}\left(\frac{\Gamma}{4 },1\right)}\right)=0,
\eeq 
with the last line being valid because\footnote{Eq. (28) in \cite{Rudnicki2015PRA} reads: $
\frac{z}{2\pi} R^2_{00}\left(z/4,1\right)\sim 1-2\sqrt{\pi z}e^{-z/2}$. This result is based on the appropriate asymptotic expansion \cite{FUCHS1964317} valid for $z\rightarrow \infty$.} $\lim_{x\rightarrow \infty}(2x/\pi)R_{00}^2(x,1)=1$. 
This results show that the coarse-grained  UR in Eq.\eqref{CGRUR} is non-trivially 
satisfied for an arbitrary (even very large) values of the coarse-graining widths. 
However, this desired property is not enjoyed by the UR
\beq
H_\alpha[P_{\Delta}^{(u)}]+H_\beta[P_{\delta}^{(v)}]\geq \ln\left(\frac{\pi  }{\alpha^{\frac{1}{2-2\alpha}}\beta^{\frac{1}{2-2\beta}}\Gamma}\right),
\label{CGRURbialy}
\eeq
first derived in \cite{bialynicki06}. This UR corresponds to Eq.\eqref{CGRUR}
in the coarse-grained regime $\Gamma/4\lesssim 1.79
$ in which $\varepsilon_1(\Gamma/4)=1/e$. Obviously, this is not a mere coincidence, as Eq. \eqref{CGRUR} subsumes \eqref{CGRURbialy}.  This is clearly visible inside the definition of $\varepsilon$ which involves the minimum of two different bounds. When $\Gamma/4> 1.79
$ the lower bound in Eq.\eqref{CGRURbialy} is negative so this UR is trivially satisfied, since the discrete entropy is always non-negative.
\par
\par
From the above considerations we can obtain an UR for the variances, $\sigma^2_{Q_{\Delta,u}}$ and $\sigma^2_{Q_{\delta,v}}$ , if we set 
$\alpha=1$ in Eq.\eqref{generalresult} and use the inequality \eqref{prodvarlessentro}:
\beq
\label{generalresultES}
\ln(2\pi e \sigma_{Q_{\Delta,u}}\sigma_{Q_{\delta,v}})\geq h[Q_{\Delta,u}]+h[Q_{\delta,v}]\geq \ln\left(\frac{\pi \hbar |\gamma|\;e^{h[D_\Delta]-\ln\Delta+h[D_\delta]-\ln\delta}}{\varepsilon_1(\Gamma/4)}\right),
\eeq
where $h[\cdot]$ stands for the Shannon entropy. Now, we can use the decompositions:
\beq
 \sigma_{Q_{\Delta,u}}^2=\sigma^2_{P_\Delta^{(u)}}+\sigma^2_{D_\Delta}\;\;\;\mbox{and}\;\;\;\sigma_{Q_{\delta,v}}^2=\sigma^2_{P_\delta^{(v)}}+\sigma^2_{D_\delta},
\eeq
where the variances of the discrete probability distributions were defined in  Eq.\eqref{DiscreteVar}, while $\sigma^2_{D_\Delta}$ and $\sigma^2_{D_\delta}$, 
are the variances of the generic HFs. Therefore, applying the above splitting to Eq.\eqref{generalresultES} we arrive at the lower bound \cite{rudnicki12b}:
\beq
\label{varianceCGURgeneral}
\left(\sigma^2_{P_\Delta^{(u)}}+\sigma^2_{D_\Delta}\right)\left(\sigma^2_{P_\delta^{(v)}}+\sigma^2_{D_\delta}\right)\geq \frac{\hbar^2 \gamma^2}{4}
\frac{e^{2(h[D_\Delta]-\ln\Delta+h[D_\delta]-\ln\delta-1)}}{\varepsilon^2_1(\Gamma/4|\gamma|)}.
\eeq
When the HF are rectangular, and in the coarse-grained regime $\Gamma/(4|\gamma|)\lesssim 1.79
$ where $\varepsilon_1(\Gamma/4|\gamma|)=1/e$, we recover the 
UR \cite{rudnicki12a}:
\beq
\label{varianceCGURrectangular}
\left(\sigma^2_{P_\Delta^{(u)}}+\frac{\Delta^2}{12}\right)
\left(\sigma^2_{P_\delta^{(v)}}+\frac{\delta^2}{12}\right)
\geq \frac{\hbar^2 \gamma^2}{4},
\eeq
where we have used the fact that in this case
\beq
\sigma^2_{D^R_\Delta}=\frac{\Delta^2}{12}\;\;\;\mbox{and}\;\;\;
\sigma^2_{D^R_\delta}=\frac{\delta^2}{12}.
\eeq
Both \eqref{varianceCGURgeneral} and \eqref{varianceCGURrectangular}
are the coarse-grained versions of the {\it Heisenberg} UR in Eq.\eqref{VarianceUR}.
It is important to emphasize that \eqref{varianceCGURrectangular} 
cannot be obtained by the simple substitution 
$\sigma^2_{P_u}\rightarrow\sigma^2_{P_\Delta^{(u)}}$ and 
$\sigma^2_{P_v}\rightarrow\sigma^2_{P_\delta^{(v)}}$ done inside the  {\it Heisenberg} UR.
\par
Although both $\sigma^2_{D_\Delta}$ and $\sigma^2_{D_\delta}$  
are the variances of a generic HF, {\it viz.} $D_\Delta(u,u_k)$ and $D_\delta(v,v_k)$ for any value of $k$, it is interesting to associate them to the respective
central bins, namely those that contain the mean value of the probability distributions $P_u$ and $P_v$. By doing this, together choosing the origins of the coordinates in the middle of the central bin, we can see that the variances $\sigma^2_{P_\Delta^{(u)}}$ and $\sigma^2_{P_\delta^{(v)}}$ are free from contributions 
associated with the statistics relevant for the central bins. Thus, if the widths of the coarse graining increase in the measurement of 
$\hat u$ and $\hat v$, the respective central bin-widths grow, so that  
the variances $\sigma^2_{P_\Delta^{(u)}}$ and $\sigma^2_{P_\delta^{(v)}}$ 
only involve contributions from the tails of the probability distributions 
$Q_{\Delta,u}$ and $Q_{\delta,v}$. Therefore, for large coarse grainings, the variances $\sigma^2_{D_\Delta}$ and $\sigma^2_{D_\delta}$ become more important in the inequalities \eqref{varianceCGURgeneral} and \eqref{varianceCGURrectangular}. Thus, in the regime when:
\bea
\Gamma&\ge& 
\pi e \Rightarrow
\Gamma\ge \frac{\pi}{\varepsilon_1(\Gamma/4|\gamma|)}\Rightarrow
\Gamma^2\ge\frac{1}{4}\frac{e^{2(h[D_\Delta]+h[D_\delta])}}
{e^2\sigma^2_{D_\Delta}\sigma^2_{D_\delta}\varepsilon^2_1(\Gamma/4|\gamma|)}\Rightarrow\nonumber\\
&\Rightarrow& \sigma^2_{D_\Delta}\sigma^2_{D_\delta}\ge
\frac{\hbar^2 |\gamma|^2}{4}
\frac{e^{2(h[D_\Delta]-\ln\Delta+h[D_\delta]-\ln\delta-1)}}{\varepsilon^2_1(\Gamma/4|\gamma|)}
\label{regimetrivially}
\eea
both \eqref{varianceCGURgeneral} and \eqref{varianceCGURrectangular} are satisfied trivially.
Note, that in Eq.\eqref{regimetrivially} we have used the relation
$4\pi^2\ge e^{2(h[D_\Delta]+h[D_\delta])}/e^2\sigma^2_{D_\Delta}\sigma^2_{D_\delta}> 0$ which can be obtained from the inequality in Eq.\eqref{prodvarlessentro}.

However,  Eq.\eqref{varianceCGURgeneral} is only the starting point for the second construction, proposed in \cite{rudnicki12b}, that is free from the above limitation, and cannot be trivially satisfied.  
This improved UR reads:
\beq
\label{CGUR-var-not-trivial}
K\left(\frac{\sigma^2_{P_\Delta^{(u)}}}{\Delta^2}\right)
K\left(\frac{\sigma^2_{P_\delta^{(v)}}}{\delta^2}\right)\geq 
\frac{\pi^2}{\Gamma^2 
\varepsilon^2_1(\Gamma/4)},
\eeq 
where $K(t)$ is implicitly defined as
\[
K(t):=\frac{\exp[2t{\cal M}^{-1}(t)]}{\erf^2\left(\sqrt{{\cal M}^{-1}(t)}/2\right)},
\]
with $\erf(x):=(2/\sqrt{\pi})\int_0^{x}e^{-y^2} dy $ being the error function and 
${\cal M}^{-1}(t)$ denoting the inverse of the invertible function
\[
{\cal M}(y):=\frac{\exp(-y/4)}{2\sqrt{\pi t}\erf(\sqrt{t}/2)}.
\]
\par
The idea behind derivation of the coarse-grained UR in Eq.(\ref{CGUR-var-not-trivial})
is the following.
Let us rewrite Eq.(\ref{varianceCGURgeneral}) in the form:
\[
\xi(h[D_\Delta],\sigma^2_{D_\Delta},h[D_\delta],\sigma^2_{D_\delta}):=\frac{\left(\sigma^2_{P_\Delta^{(u)}}+\sigma^2_{D_\Delta}\right)\left(\sigma^2_{P_\delta^{(v)}}+\sigma^2_{D_\delta}\right)}
{e^{2(h[D_\Delta]+h[D_\delta]-1)}}
\geq \frac{1}{4\Gamma^2\varepsilon^2_1(\Gamma/4)}.
\]
Now the function $\xi$
is supposed to be minimized, however, because the Shannon entropy $h[D_\Delta]$ ($h[D_\delta]$)
is interrelated with (bounded by a function of) the variance $\sigma^2_{D_\Delta}$ ($\sigma^2_{D_\delta}$)
the minimization needs to be performed in two steps.
For fixed values of the variances $\sigma^2_{D_\Delta}$ and $\sigma^2_{D_\delta}$, the function $\xi$ achieves 
its minimum when  
the Shannon entropies $h[D_\Delta]$  and $h[D_\delta]$ are maximized with respect to 
the functional form of the HFs, $D_\Delta$ and $D_\delta$. As already stated, the HFs are constrained by the requirement of the fixed value for both variance. The form of the HF with maximum 
Shannon entropy \cite{rudnicki12b} is a Gaussian with support restricted to the central bin and whose variance is an appropriate function\footnote{For details see \cite{rudnicki12b}.} of $\sigma^2_{D_\Delta}$ ($\sigma^2_{D_\delta}$). Therefore, for this optimal HF its Shannon entropy $h[D^{op}_\Delta]$ ($h[D^{op}_\delta]$) is only a function of the variance $\sigma^2_{D_\delta}$ ($\sigma^2_{D_\delta}$), thus we 
have $G(\sigma^2_{D_\Delta},\sigma^2_{D_\delta})=\xi(h[D^{op}_\Delta],\sigma^2_{D_\Delta},h[D^{op}_\delta],\sigma^2_{D_\delta})$.
The second step is a direct minimization of $G(\sigma^2_{D_\Delta},\sigma^2_{D_\delta})$, which results in the left hand side product in Eq.(\ref{CGUR-var-not-trivial}).
\par
According to the discussion above Eq.(\ref{regimetrivially}) 
the coarse-grianed UR in Eq.\eqref{CGUR-var-not-trivial} has no contributions 
from the statistics corresponding to the central bin.
In the limit when $\Delta,\delta\rightarrow 0$ we recover the {\it Heisenberg} UR 
in Eq\eqref{VarianceUR} thanks to the identities \cite{rudnicki12b}
\beq
\lim_{\Delta \rightarrow 0} \Delta^2 K\left(\frac{\sigma^2_{P_\Delta^{(u)}}}{\Delta^2}\right)
=\sigma^2_{P_u} \lim_{y\rightarrow0}
\frac{1}{{\cal M}(y)}\frac{\exp(2y{\cal M}(y))}{\erf^2\left({\sqrt{y}/2}\right)}=
2\pi e \sigma^2_{P_u}.
\eeq
In the opposite limit of infinite coarse graining, {\it viz} $\Delta,\delta\rightarrow \infty$, we have $\sigma_{P_\Delta^{(u)}}^2,\sigma_{P_\delta^{(v)}}^2\longrightarrow 0$
and
\beq
\overbrace{\lim_{\sigma_{P_\Delta^{(u)}}^2 \rightarrow 0} K\left(\frac{\sigma^2_{P_\Delta^{(u)}}}{\Delta^2}\right)}^{=1}
\overbrace{\lim_{\sigma_{P_\delta^{(v)}}^2 \rightarrow 0} K\left(\frac{\sigma^2_{P_\delta^{(v)}}}{\delta^2}\right)}^{=1}\ge
\overbrace{\lim_{\Gamma \rightarrow \infty}\frac{\pi^2}{\Gamma^2
\varepsilon^2_1(\Gamma/4)}}^{=1}.
\label{limit-ones}
\eeq
It is important to note that since
\beq
\frac{\pi^2}{\Gamma^2 
\varepsilon^2_1(\Gamma/4)}>1,
\eeq
whenever both $\Delta$ and $\delta$ are finite, it is forbidden to set $\sigma_{P_\Delta^{(u)}}^2$
 and $\sigma_{P_\delta^{(v)}}^2$ as simultaneously equal to zero, as it would contradict 
 the coarse-grained UR \eqref{CGUR-var-not-trivial}. This means 
 that any quantum state (pure or mixed) cannot be localised in both observables 
 $\hat u$ and $\hat v$ that are CCOs. In other words, the associated probability distributions
cannot simultaneously have compact support. 


This remarkable conclusion somehow threatens the scientific program to recover classical mechanics solely from  coarse-grained averaging, physically originating from the finite-precision of the observations \cite{ballentine98,Kofler2007a,Kofler2007}.
Indeed, quantum features can be observed in the measurement 
of $\hat u$ and $\hat v$ 
irrespective of the precision of the detectors. 
 However, for very large coarse-graining widths the variances $\sigma_{P_\Delta^{(u)}}^2$
 and $\sigma_{P_\delta^{(v)}}^2$ are dominated by the contributions 
 from  the tails of the $P_\Delta^{(u)}$ and  $P_\delta^{(v)}$. Thus, as these probabilities are likely very small, they would be particularly susceptible to statistical fluctuations and it would in general require very long acquisition times to collect the sufficient amount of data necessary to verify the UR \eqref{CGUR-var-not-trivial} in the regime of extremely large coarse graining.
\subsection{URs valid for general observables, $\hat u$ and $\hat v$, defined in Eq.\eqref{defuv}. }
\par
If we let $\alpha=1$ in Eq.\eqref{generalresult}, use rectangular HFs such that Eq.\eqref{entropies-rec-HF} is valid and restrict the size of the involved bins such that $\varepsilon_1(\Gamma/4|\gamma|)=1/e$ --- this is the regime of the coarse graining when $\Gamma/4\lesssim 1.79$ --- we obtain the simplified coarse-grained UR of the form:
\beq
\label{entropic-CG-UR-for-general-u-v}
h[Q_{\Delta,u}]+h[Q_{\delta,v}]\geq \ln\left(\pi e \hbar  |\gamma|\right).
\eeq
Because the coarse-grained UR
in Eq.\eqref{generalresult} was derived only for CCOs, $\hat u$ and $\hat v$, {\it a priori}
it is not clear why the above UR could remain
valid also for generalized observables defined in Eq.\eqref{defuv}. This fact, however, can be proved 
with the help of the {\it Shannon-entropy} UR \eqref{ShannonUR}, that has properly been extended to the desired observables, and the inequalities:
\beq
\label{hQgreaterhP}
h[Q_{\Delta,u}]\geq
h[P_u]
\;\;\;\mbox{and}\;\;\;
h[Q_{\delta,v}]\geq
h[P_v],
\eeq
whose detailed derivation based on the Jensen inequality is relegated to Appendix \ref{AppendixB}.
Passing to the discrete entropies we find the coarse-grained UR:
\beq
H[P_{\Delta}^{(u)}]+H[P_{\delta}^{(v)}]\geq \ln\left(\frac{\pi e   }{\Gamma}\right),
\label{CGRURbialy-shannon}
\eeq 
which looks the same as the one derived in \cite{bialynicki84} for CCOs. Here, the validity of this UR has been extended for any observables $\hat u$ and $\hat v$ as defined in Eq.\eqref{defuv}. 
Also, following the same arguments that lead from Eq.\eqref{generalresultES}
to  the UR in Eq.\eqref{varianceCGURrectangular} we can see that the UR for the discrete variances is also valid for 
general $\hat u$ and $\hat v$ as defined in Eq.\eqref{defuv}.
\par
To briefly summarize, entropic uncertainty relations for coarse-grained probability distributions
were almost only considered for position and momentum variables. As far as we know, the only exceptions are given in Refs. \cite{huang11,guanlei09}. However, as we have shown here, the generalization of entropic 
URs for differential probabilities associated with  general observables
$\hat u$ and $\hat v$, which are linear combinations of position and momentum, can be done in many cases.  However, in each 
case a careful analysis  should be carried out to verify that the 
related coarse-grained URs are also valid for these generalised operators.
Here, we have done this only in the simple cases.

\subsection{Coarse-grained URs merged with the majorization approach }
In \cite{Rudnicki2015PRA}  the coarse-grained scenario has been discussed with the help of 
the results obtained in \cite{friedland2013,puchala2013,RPZ2014}, namely the majorization-based approach to quantification of uncertainty. To say it briefly, a majorization relation $x\prec y$ between two arbitrary $d$-dimensional probability distributions means that for every $n\leq d$ the inequality $\sum_{k=1}^{n}x_{k}^{\downarrow}\leq\sum_{k=1}^{n}y_{k}^{\downarrow}$ holds, with an equality (normalization) for $n=d$. Traditionally, by “$\downarrow$” we denote the decreasing order, so that $\bigl(x^{\downarrow}\bigr){}_{k}\geq\bigl(x^{\downarrow}\bigr){}_{l}$, for all $k\leq l$. The Rényi entropy (and also others, such as the Tsallis entropy) is Schur-concave, which implies $H_{\alpha}\left[x\right]\geq H_{\alpha}\left[y\right]$ whenever $x\prec y$.

In the context of coarse-grained
probability distributions it was conceptually simpler to consider
the so-called direct-sum majorization introduced in \cite{RPZ2014}. An advantage of the majorization approach is that it covers a regime of ($\alpha,\beta$) parameters, $\beta=\alpha$ to be precise, which in some way is perpendicular to the conjugate choice $1/\alpha+1/\beta=2$. In \cite{Rudnicki2015PRA} an
infinite hierarchy of majorization vectors, depending on a single parameter
$\Gamma=\Delta\delta/\hbar$, has been derived. The discussion is conducted for CCOs, thus one can easily recognize the dimensionless $\Gamma$ parameter as those which appears in all previous URs \fabri{with $\gamma=1$}.

The main result, namely
a family of lower bounds denoted as $\mathcal{R}_{\alpha}^{(n)}\left(\Delta\delta/\hbar\right)$ for $n=2,\dots,\infty$,
has been presented in Eq. (27) from \cite{Rudnicki2015PRA}, however, we refrain from providing its detailed construction here. It seems enough to say that the bound in question is a function of $R^2_{00}\left(j_0 \Gamma /4,1\right)$ with $j_0$ being certain positive integers. In other words, in spirit, the majorization bound is close to that derived in \cite{rudnicki12b} and extensively discussed above. A comparison of the new bound and \eqref{CGRUR} for $\alpha=1=\beta$ --- the only value of both parameters for which the involved bounds describe the same situation --- showed that $\mathcal{R}_{1}^{(3)}$ outperforms \eqref{CGRUR} in the regime when the $R_{00}$-term does contribute to $\varepsilon_1$.

Asymptotic behavior of the new and previous coarse-grained bounds
shows that for $\alpha=1=\beta$ and large $\Gamma$, all $\mathcal{R}_{1}^{(n)}$ bounds
improve (\ref{CGRUR}) by a divergent factor $\Gamma/4$. Moreover, the typical behavior of discrete majorization bounds has been confirmed
in the coarse-grained setting. In the discrete case, the majorization
relations almost surely dominate the MU bound, with an
exception being a small neighborhood of the point for which the unitary
matrix $\mathbf{U}$ is the Fourier matrix. The analog of the Fourier matrix
in the coarse-grained scenario is the continuous limit $\Gamma\rightarrow0$.
This probably intuitive fact has been rigorously shown by means of
the asymptotics of $\mathcal{R}_{1}^{(\infty)}$ for small $\Gamma$,
which is equal to $-\frac{1}{2}\ln\Gamma$.

\subsection{Other coarse-grained URs}
At the very end of this long section we would like to touch upon few coarse-grained URs which go beyond the standard position-momentum conjugate pair. First of all, Bialynicki-Birula also provided his major {\it Shannon entropy} UR in the case of angle and angular momentum \cite{bialynicki84}, as well as\footnote{Together with Madajczyk.} to the variables on the sphere \cite{madajczyk}. Coarse graining in these physical settings is only relevant for the periodic CVs (angle on a circle and two angles on a sphere), as the conjugate variables are discrete (though infinite dimensional).

Also, the coarse-grained scenario has been developed \cite{FurrerMem} in relation to the memory-assisted UR \cite{berta10} relevant for quantum key distribution. The result, even though non-trivial, differs from Eq. \eqref{CGRUR} in a similar fashion as the MU bound differs from the UR in the presence of quantum memory by Berta et al \cite{berta10}.   

Going in a completely different direction, Rastegin \cite{RASTEGIN17} in his recent contribution proposed an extension of \eqref{CGRURbialy} to the case of a modified CCR, which assumes the form $[\hat x,\hat p]=i\hbar(1+\beta \hat p^2)$. The parameter\footnote{Not to be confused with $\beta$ playing the role of a conjugate parameter in the MU bound and similar URs for the R{\'e}nyi entropies.} $\beta$ is related to the so-called minimal length predicted by certain variants of string theory and similar approaches. 

Last but not least, some of us have very recently derived an inequality (see Eqs. 9-12 from \cite{tasca18b}), which could be understood as an UR (valid for CCOs) in the setting relevant for periodic coarse graining discussed in section \ref{PCGsect}. As this UR involves additional averaging of $\rom p^{(x)}_k (\hat{\rho})$ and  $\rom p^{(p)}_l (\hat{\rho})$ defined below Eq. \eqref{MUB_CG} with respect to the positioning degrees of freedom,  we do not provide further details of this construction encouraging the interested reader to consult \cite{tasca18b}.

\section{Applications of coarse-grained measurements and coarse-grained Uncertainty Relations}
\label{sec:appCGUR}
{As discussed above, when detecting the position and momentum of particles such as photons or individual atoms, coarse-grained measurements are not just necessary but can be much more practical.  In this regard, URs that deal with coarse-grained measurements can be useful for a number of applications, such as those discussed in section \ref{sec:relevance}.
\par
In Ref. \cite{schneeloch13} EPR-steering was tested for discrete distributions of measurements made from standardized binning on the two-photon state produced from spontaneous parametric down-conversion, using a coarse-grained version of the EPR-steering criteria of Ref. \cite{walborn11a}.  Bi-dimensional steering was observed for sample sizes ranging from $8 \times 8$ to $24 \times 24$, representing a considerable reduction in measurement overhead when compared with the quasi-continuous measurements reported in Ref. \cite{walborn11a}, which sampled about 100 data points per cartesian direction (about $10^4$ total measurements) to evaluate entropic EPR-steering criteria of continuous variables.      
\par
The pitfalls of applying the usual entanglement criteria for continuous variables to coarse-grained measurements was shown in Ref. \cite{Tasca13}, where it was demonstrated that this can lead to false-positive identifications of entanglement, such that the separability criteria based on uncertainty relations discussed in section \ref{sec:relevance} can be (falsely) violated even for separable states.   
 To show how binned data should be properly handled, a coarse-grained UR was used, along with the PPT argument to properly identify entanglement experimentally in a system of spatially-entangled photons.   In particular, a variance criteria based on \eqref{varianceCGURrectangular} was tested for the global operators defined in equations \eqref{eq:upm} and \eqref{eq:vpm}.    
It was also shown that coarse-grained entropic entanglement criteria, for example based on inequality \eqref{generalresult} ($\alpha=\beta=1$) applied to operators  \eqref{eq:upm} and \eqref{eq:vpm}, can be superior to coarse-grained variance based criteria, identifying entanglement when variance criteria do not, even for the case of Gaussian states.  This is due to the fact that the coarse-grained probability distributions functions such as those shown in figure \ref{Fig:CGdistributions} are not Gaussian functions, even when the quantum state under investigation is Gaussian. 
\par
Standard coarse graining has been studied in the context of quantum state reconstruction of single and two-mode Gaussian states, and the quantum to classical transition \cite{park14}.  Two scenarios were considered: direct reconstruction of the covariance matrix alone, and full reconstruction of the state using maximum likelihood estimation.  The reconstructed coarse-grained functions were compared to those of Gaussian states subject to thermal squeezed reservoirs, indicating that in this context coarse graining does not produce a thermalized (decohered) Gaussian state.
\par
The work mentioned above considered standard coarse graining, as described in section \ref{sec:CGmodels}.  In some cases it is interesting to consider different models, such as that of periodic coarse graining described in section \ref{PCGsect}. The mutual unbiasedness of periodic coarse graining described in section \ref{MUBsect} has been tested experimentally for two \cite{tasca18a} and even three \cite{paul18a} phase-space directions.  It was shown that mutual unbiasedness appears when the appropriate bin widths of the two or three conjugate variables are chosen.    
Periodic coarse graining has also been used in the detection of spatial correlations of photon pairs from SPDC \cite{tasca18b}.  Using a novel entanglement criteria based on the UR for characteristic functions \cite{rudnicki16}, it was possible to identify entanglement with as few as $2 \times 2$ measurements in position and momentum (8 total), representing a considerable reduction in measurement overhead.    
\par
Simple binary binning of homodyne measurements has been proposed as a means to test dichotomic Bell's inequalities in CV systems, while allowing for high detection efficency \cite{gilchrist98,gilchrist99,munro99,garciapatron04}.    Other types of non-standard coarse graining have been proposed as a means to violate Bell's inequality using homodyne measurements on non-Gaussian states \cite{wenger03}.  Though it was shown that one could achieve maximal violation in principle, exotic non-Gaussian states are required.    In Ref. \cite{tasca09b} it was shown that imperfect binning could result in false violations of Bell's inequalities, and even in violations of Cirelson's bound for quantum Bell correlations.
\par
A closely related subject to periodic coarse graining of CVs is that of the so called modular variables \cite{aharanov,vernazgris14,ketterer16}, for which phase-space variables $u$ are rewritten as $u=n_u \ell + \bar{u}$, where $n_u$ is the integer component and $\bar{u}$ the modular component, such that $0 \leq \bar{u} < \ell$.  Here $\ell$ is a scaling parameter of appropriate dimension.  For two CCOs, such as $\oper{x}$ and $\oper{p}$ for example, the integer operator of one observable--say--$\oper{n}_x$ and the modular operator of the other observable--$\oper{\bar{p}}$ satisfy URs that closely resemble those of the angle and angular momentum variables \cite{bialynicki84}.             
The modular variable construction was first introduced by Aharanov et al. \cite{aharanov69,aharanov} as a method to identify non-locality in quantum mechanics.  Since then, a number of interesting applications have been developed.   Variance-based URs for the modular variable construction were proposed as a method to identify a novel type of squeezing, as well as entanglement in pairs of atoms \cite{gneiting11}.  This entanglement criteria was used in Refs. \cite{carvalho12}, along with one based on entropic uncertainty relations, to identify spatial entanglement of photon pairs that have passed through multiple slit apertures.  Application to multiple-photon states was studied in Ref. \cite{barros15}.   It is worth noting that in this case the usual CV entanglement criteria as discussed in section \ref{sec:relevance} are incapable of detecting entanglement.   
Modular variables have been proposed as a way to test for the Greenberger-Horne-Zeilinger paradox in CV systems \cite{massar01}, as well as quantum contextuality \cite{plastino10,asadian15,finot17} and as a method to construct algebras resembling that of discrete systems \cite{vernazgris14,asadian16,ketterer16}.      
}

\section{Conclusion}
\label{sec:conc}
Uncertainty relations play an important role in quantum physics, which is two-fold:  on the one hand they have historically represented the difference between classical and quantum physics, while on the other hand they are a tool that can be used to identify and even quantify interesting quantum properties.  Beginning with the seminal work of Heisenberg in 1927, a number of uncertainty relations have been developed for continuous variable quantum systems.  However, in a realistic experimental setting, one never has access to the infinite dimensional spectrum associated to these observables.  Thus, coarse graining is imposed by the detection apparatus to account for the measurement precision and range.  
\par
Here we have provided a review of a number of quantum mechanical uncertainty relations tailored specifically to coarse-grained measurement of continuous quantum observables.  Our aim was to survey the state-of-the-art of the subject, from both the theoretical advances to experimental application of coarse-grained uncertainty relations.  
\par
A number of interesting open questions  remain. First, it would be interesting to see
the generalization of all the coarse-grained URs presented here for 
pairs of observables that are connected by general unitary metaplectic transformations.  Second, one can consider applying coarse graining to URs not mentioned explicitely here, such as the triple variance product criteria \cite{Weigert08,paul16}, UR for characteristic functions \cite{rudnicki16}, among others, as well the plethora of moment inequalities arising from tests for non-classicality \cite{shchukin05c,ryl15} and entanglement \cite{shchukin05,agarwal05,hillery06a}. 
Third, and more important, a deep discussion of the role of coarse-grained URs 
within the scientific program to recover classical mechanics solely from coarse-grained averaging should be developed. We hope that this review encourage this discussion.

\section{Conclusions}





\acknowledgments{The authors acknowledge
financial support from the Brazilian funding agencies CNPq, CAPES (PROCAD2013 project)
and FAPERJ, and the National Institute of Science and Technology -
Quantum Information. \L .R. acknowledges financial support by grant
number 2015/18/A/ST2/00274 of the National Science Center, Poland.}

\authorcontributions{All authors contributed equally to this work.}

\conflictsofinterest{The authors declare no conflict of interest. The founding sponsors had no role in the design of the study; in the collection, analyses, or interpretation of data; in the writing of the manuscript, and in the decision to publish the results.} 

\abbreviations{The following abbreviations are used in this manuscript:\\

\noindent 
\begin{tabular}{@{}ll}
CV & Continuous variable\\
UR & Uncertainty relation\\
QIT & Quantum information theory\\
CCR & Canonical commutation relation\\
CCO & Canonically conjugate operators\\
pdf & probability distribution function\\
EPR & Einstein-Podolsky-Rosen\\
PPT & Positive partial transposition\\
PCG & Periodic coarse graining\\
MU & Maassen-Uffink\\
HF & Histogram function\\
\end{tabular}}

\appendixtitles{no} 
\appendixsections{multiple} 
\appendix
\section{}
\label{AppendixA}
Following \cite{rudnicki12b} we aim to prove the decomposition in Eq.\eqref{decomdiffRenyi}. To this end it is enough to discuss the case of $\hat u$ since the proof for $\hat v$ looks the same. We can write:
\bea
h_\alpha[Q_{\Delta,u}]&=&\frac{1}{1-\alpha}\ln\left(\int_{-\infty}^\infty du\; [Q_{\Delta,u}(u)]^\alpha\right)=
\frac{1}{1-\alpha}\ln\left(\sum_{k\in {\mathcal Z}_k}\int_{{\mathcal R}_k} du\; [Q_{\Delta,u}(u)]^\alpha\right)\nonumber\\
&=&\frac{1}{1-\alpha}\ln\left(\sum_{k\in {\mathcal Z}_k} \left[p^{(u)}_{\Delta,k}\right]^\alpha\int_{{\mathcal R}_k} du\;[D_\Delta(u,u_k)]^\alpha\right),
\eea
where we use the fact that the function $Q_{\Delta,u}(u)$ in the interval $\mathcal R_k$ is equal to
$ \rom p^{(u)}_{\Delta,k}\;D_\Delta(u,u_k)$. Now, because the shape of the HF, $D_\Delta(u,u_k)$, is the same for all values of $k$, the integral $\int_{{\mathcal R}_k} du\; [D_\Delta(u,u_k)]^\alpha$ does not depend on $k$.
Therefore, we can write:
\beq
h_\alpha[Q_{\Delta,u}]=\frac{1}{1-\alpha}\ln\left(\sum_{k\in {\mathcal Z}_k} \left[p^{(u)}_{\Delta,k}\right]^\alpha\right)+
\frac{1}{1-\alpha}\ln\left(\int_{{\mathcal R}_k} du\; [D_\Delta(u,u_k)]^\alpha\right),
\eeq
that corresponds to the decomposition in Eq.\eqref{decomdiffRenyi}.
\section{}
\label{AppendixB}
Here, we prove the inequalities in Eq.\eqref{hQgreaterhP}. As before, is enough to consider the single case relevant for the variable $u$.
What we do in the next few lines, we actually closely follow the treatment presented in \cite{bialynicki11}. First we define the mean value
within the $k$th histogram bin:
\beq
\expval{\ldots}_k:=\frac{1}{\Delta}\int_{(k-1/2)\Delta}^{(k+1/2)\Delta}\ldots du.
\eeq
Then, because the function $f(x)=x\ln(x)$ is convex we can apply Jensen's 
inequality \cite{cover} to obtain,
\beq
\expval{P_u\ln P_u}_k\geq \expval{P_u}_k\ln\expval{P_u}_k.
\eeq
Now we can use the definition in Eq.\eqref{dyskretne2}, multiply both sides 
by $-1$ and sum over $k\in {\cal Z}_k$:
\beq
-\sum_{k\in {\cal Z}_k}\rom p_{\Delta,k}^{(u)}\ln\rom p_{\Delta,k}^{(u)}+\left(\sum_{k\in {\cal Z}_k}\rom p_{\Delta,k}^{(u)}\right)\ln(\Delta)\ge
-\sum_{k\in {\cal Z}_k}\int_{(k-1/2)\Delta}^{(k+1/2)\Delta}P_u(u)\ln P_u(u) \;du.
\eeq 
After using the condition in Eq.\eqref{sum-prob-approx-1}, the definition of the 
discrete Shannon entropy $H[P_\Delta^{(u)}]:=-\sum_{k\in {\cal Z}_k}\rom p_{\Delta,k}^{(u)}\ln\rom p_{\Delta,k}^{(u)}$, the decomposition in Eq.\eqref{decomdiffRenyi} with $\alpha=1$ and $h[D^R_\Delta]=\ln\Delta$, and the definition of the 
differential Shannon entropy in Eq.\eqref{defShannonEntropy} we obtain:
\beq
h[Q_{\Delta,u}]:=H[P_\Delta^{(u)}]+\ln\Delta\ge h[P_u], 
\eeq
which is the desired result.

\end{document}